\documentclass[acmtog, nonacm]{acmart}

\acmSubmissionID{296}

\usepackage{booktabs} %

\usepackage[ruled]{algorithm2e} %

\SetAlFnt{\small}
\SetAlCapFnt{\small}
\SetAlCapNameFnt{\small}
\SetAlCapHSkip{0pt}

\authorsaddresses{
\\This is the author's version of the work. It is posted here for your personal use. Not for redistribution. }

\newcommand{\sysName}{Sketch3DVE}

\usepackage{color} 

\newcommand{\rv}[1] {{\color{black}  #1}}

\newcommand{\hb}[1]{{\color{black}            {#1}}}

\newcommand{\lfl}[1]  {{\color{black} #1}}

\begin{document}
\title{\sysName: Sketch-based 3D-Aware Scene Video Editing}

\author{Feng-Lin Liu} 
\affiliation{
\institution{Institute of Computing Technology, Chinese Academy of Sciences, China and University of Chinese Academy of Sciences}
\country{China}
}
\email{liufenglin21s@ict.ac.cn}

\author{Shi-Yang Li} 
\affiliation{
\institution{Institute of Computing Technology, Chinese Academy of Sciences}
\country{China}
}
\email{shiyangli179@gmail.com}

\author{Yan-Pei Cao} 
\affiliation{
\institution{VAST}
\country{China}
}
\email{caoyanpei@gmail.com}

\author{Hongbo Fu}
\affiliation{
\institution{Hong Kong University of Science and Technology}
\country{China}
}
\email{hongbofu@ust.hk}

\author{Lin Gao}
\authornote{Corresponding author is Lin Gao (gaolin@ict.ac.cn).}
\affiliation{
\institution{Institute of Computing Technology, Chinese Academy of Sciences, China and University of Chinese Academy of Sciences }
\country{China}
}
\email{gaolin@ict.ac.cn}

\renewcommand\shortauthors{Feng-Lin Liu, Shi-Yang Li, Yan-Pei Cao, Hongbo Fu, and Lin Gao}

\begin{teaserfigure}
  \centering
  \includegraphics[width=1.0\linewidth]{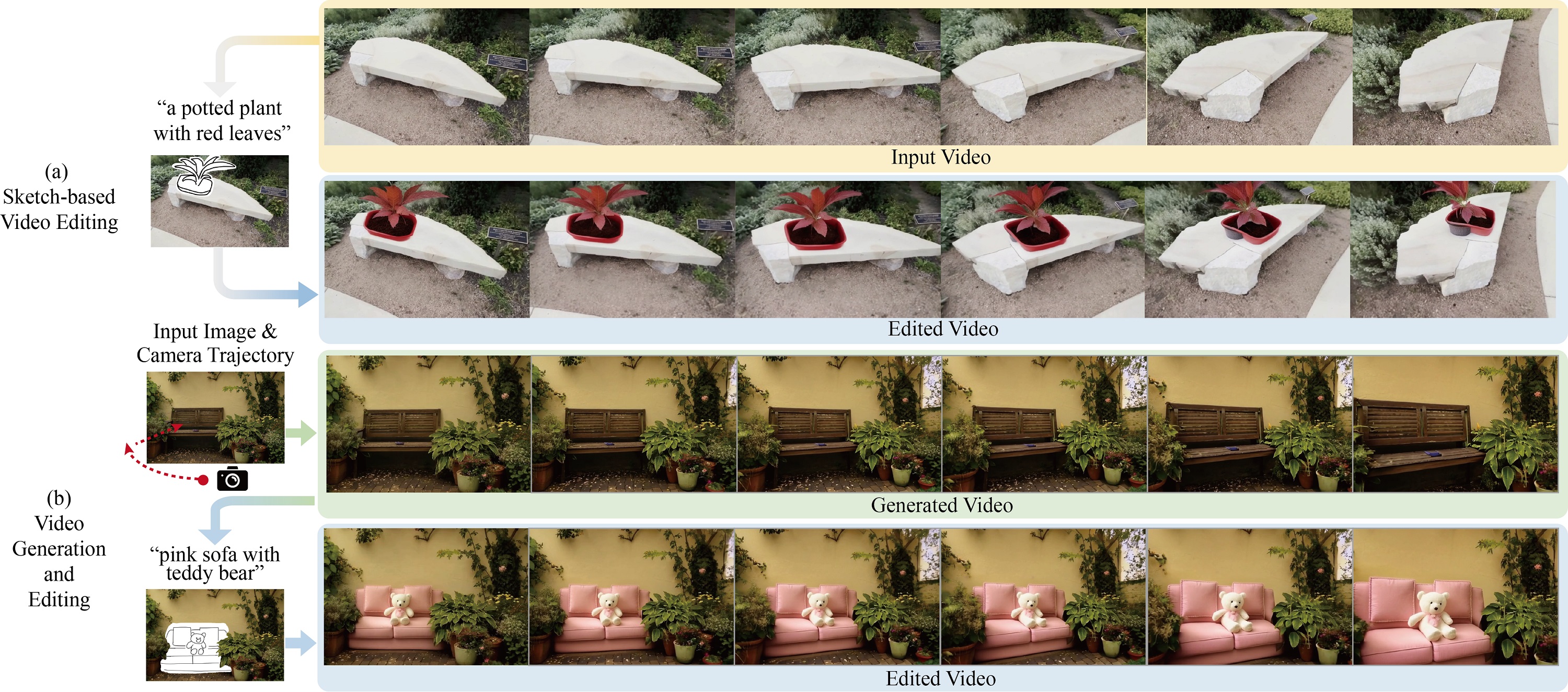}
  \caption{Given an input video (yellow box in (a)) with significant viewpoint changes, {\sysName} generates realistic editing results (blue box) with the inputs of {a text prompt,} %
  hand-drawn sketch, and {a mask (in white)}. %
  Our method can also edit a video generated given an input image and a camera trajectory (b). Input Video ©DL3DV-10K, Input image ©Pegah Sharifi.
  }
  \label{fig:teaser}
\end{teaserfigure}

\begin{abstract}

Recent video editing methods achieve attractive results in style transfer or appearance modification.
However, editing the structural content of 3D scenes in videos remains challenging, particularly when dealing with significant viewpoint changes, such as large camera rotations or zooms. 
Key challenges include generating novel view content that remains consistent with the original video, preserving unedited regions, and translating sparse 2D inputs into realistic 3D video outputs.
To address these issues, we propose {\sysName}, a sketch-based 3D-aware video editing method to enable detailed local manipulation of videos with significant viewpoint changes.
To solve the challenge posed by sparse inputs, we employ %
image editing methods to generate edited results for the first frame, which are then propagated to the remaining frames of the video.
We utilize sketching as an interaction tool for precise geometry control, while other mask-based image editing methods are also supported.
To handle viewpoint changes, we perform a detailed analysis and manipulation of the 3D information in the video.
Specifically, we utilize a dense stereo method to estimate a point cloud and the camera parameters of the input video. 
We then propose a point cloud editing approach that uses depth maps to represent the 3D geometry of newly edited components, aligning them effectively with the original 3D scene. 
To seamlessly merge the newly edited content with the original video while preserving the features of unedited regions, we introduce a 3D-aware mask propagation strategy and employ a video diffusion model to produce realistic edited videos.
Extensive experiments demonstrate the superiority of {\sysName} in video editing. 
Homepage and code: \href{http://geometrylearning.com/Sketch3DVE/}{http://geometrylearning.com/Sketch3DVE/}

\end{abstract}



\maketitle

\section{Introduction}

Video generation and editing are popular research areas with broad applications in movie production, education, robotics, and AR/VR. 
Unlike creating videos from scratch, editing involves manipulating the geometry and appearance of objects in existing real or synthetic videos, enabling detailed user customization.
While pioneering video editing methods \cite{Stitch_it_in_Time, Layered_Video_Editing, Style_Transfer_video, Stylize_Video_example, DeepFaceVideoEditing} achieve attractive results, they are limited to specific domains or restricted editing types. 
Image editing methods, such as sketch-based works \cite{SC-FEGAN, SKED, SketchEdit, SketchDream, Faceshop} and mask-based inpainting works \cite{FramePainter, Paint_example, FLUX_inpainting}, have good flexibility and attractive results. 
However, applying similar ideas to video editing presents several challenges, including domain gaps between single-image 2D inputs and videos, the absence of novel view information for videos with significant view changes, and the requirement to identify and preserve unedited regions. 

Recent advancements in diffusion models have led to impressive results in text- and image-based video generation, with both commercial {\cite{sora, keling, stableVideo, Gen3}} and open source works \cite{cogvideox, opensora, Open-Sora-Plan, Vchitect}.
Similarly, diffusion-based video editing methods \cite{AVID, AnyV2V, I2VEdit, ReVideo} have made significant progress in diverse editing scenarios. 
However, as shown in Fig. \ref{fig:compare}, most of these methods transfer motion features from the input video to the edited video, which is effective for appearance modification but struggles with structural editing, such as component insertion and replacement.
Furthermore, these methods use UNet or DiT networks to handle temporal motion implicitly. However, they lack extensive 3D information analysis, limiting their ability to process videos with substantial viewpoint changes. 

Recent video generation methods have employed 3D information to enable controllable viewpoint changes. 
Given a single image, novel view videos can be synthesized by inserting camera parameters into pretrained video diffusion models \cite{Direct-a-Video, MotionControl, CameraCtrl, CamI2V}, or using point cloud rendering for explicit view control \cite{ViewCrafter, ViewExtrapolation}.
These methods support editing by modifying the input image and generating videos with the original camera motions. 
However, for generated novel view frames, the obscured areas and components outside the viewport of the edited image are entirely imagined by the models, often resulting in unintended changes in these regions even if they remain unedited.
Instead of video generation, NeRF \cite{NeRF} and 3DGS \cite{3DGS} have been utilized for 4D generation \cite{BAGS, CAT4D, Animatable_gaussian_human}, image-to-object \cite{zero123, image_to_voxel}, and image-to-scene \cite{GenWarp}, while focusing on generation from scratch rather than local modification. 3D scene editing methods \cite{RD-VIO, Infusion, SPIn-NeRF} modify 3D attributes to render edited videos, but struggle with structural modifications or produce rendering-style outputs.

To address these issues, we propose \sysName, a sketch-based 3D-aware method to produce realistic edited videos (see Fig. \ref{fig:teaser}).
We assume both the input and output are scene videos with significant viewpoint changes but slight object movements.
To bridge the domain gap between 2D sketches and videos, we first generate the edited results of the first frame by MagicQuill \cite{MagicQuill} and then propagate the editing effects across the video. Image editing on the first frame can also be achieved with other image editing methods (e.g., \cite{FLUX_inpainting, Paint_example}) that use masks to label the local modification regions.
To generate high-quality details in novel views %
of the edited content while ensuring seamless camera motion consistency with the input video, we perform explicit 3D geometry analysis and manipulation.
Specifically, we first obtain a point cloud and the camera parameters from the input video using DUSt3R \cite{DUSt3R}.
Then, we propose a point cloud editing approach to map 2D image edits into 3D space, ensuring that the edited regions align with the rest of the scene. 
Depth maps, which provide relative 3D geometry information and explicit pixel matching before and after editing, are utilized to achieve this alignment.
To solve the issue of unedited region identification and preservation, we propose a 3D-aware mask propagation strategy that tracks the input 2D mask across other frames. 
Finally, a video diffusion model generates the novel view results of the edited components that are merged with unedited regions to synthesize a realistic edited video.
Our method supports a wide range of editing operations, including object insertion, removal, replacement, and modification to shape and texture, as shown in Fig. \ref{fig:results} and \ref{fig:application_color}.

Extensive experiments demonstrate that our method produces higher-quality video editing results than existing image-based video generation and editing approaches.
Our main contributions can be summarized as follows:
\begin{itemize}
\item We propose a novel sketch-based 3D-aware video editing method that generates realistic and high-quality editing results for videos with significant viewpoint changes.
\item We propose a point cloud editing approach that utilizes depth maps to represent and align the edited regions with the original 3D scene. 
\item We develop a 3D-aware mask propagation strategy and a {precise region} modification video diffusion model that synthesizes novel view results of edited components while accurately preserving unedited regions. 
\end{itemize}

\section{Related Work}

Our method is closely related to camera-controllable video generation, deep video {editing, and sketch-based content editing.}

\paragraph{Camera Controllable Video Generation}
One category of methods implicitly inserts camera parameters into video generation models.
The camera features influence the temporal attention mechanism \cite{Direct-a-Video, MotionControl}, or transform to Pl{\"u}cker Embedding \cite{plucker_embedding} with additional condition networks for insertion \cite{CameraCtrl, VD3D, CPA}.
Epipolar attention \cite{CamI2V, CamCo}, {LoRa} fine-tuning \cite{DimensionX}, and multi-view dataset{s} \cite{SynCamMaster} further enhance camera control performance. 
Explicit methods, on the other hand, construct 3D representations from input images and render them into novel views, which are then used as conditions for video generation.
Training-free denoising resampling \cite{Training-free-camera,ViewExtrapolation} and video diffusion model finetuning \cite{ViewCrafter} are utilized to translate point cloud renderings into videos. 
3D tracking videos \cite{Diffusion_as_Shader} and 3D box rendering depth maps \cite{CineMaster} further serve as conditions for dynamic content synthesis.
While these methods effectively control the camera %
for video generation, they are inherently limited for editing due to the absence of relationship analysis between the original video content and the newly introduced modifications. 
Our method generates high-quality editing contents that seamlessly merge into original videos with good view consistency.

\paragraph{Deep Video Editing}
Pioneering works achieve effective video editing in facial videos \cite{Stitch_it_in_Time, DeepFaceVideoEditing}, style transfer \cite{Style_Transfer_video, Stylize_Video_example}, and layered editing propagation \cite{Layered_Video_Editing}. 
With the success of diffusion models, many works \cite{Rerender_A_Video, zero-shot-editing, FateZero, Video-P2P, CoDeF, Pix2Video} extend text-to-image generation models \cite{StableDiffusion} to video editing by enforcing temporal coherence. 
Video generation models are further used to mitigate flickering and enhance robustness.
AVID \cite{AVID} treats video editing as a sketch-guided inpainting task and enables text-based editing. 
Similar to our method, many works propagate the editing from the first frame of a video to the rest.
AnyV2V \cite{AnyV2V}, a training-free method, merges edited results with attention features of the input video but sometimes generates structural distortion and temporal inconsistency. 
I2VEdit \cite{I2VEdit} fine-tunes models using LoRa on temporal layers to extract motion information, producing consistent outputs but struggles with object insertion due to missing motion information in the original video.
ReVideo \cite{ReVideo} {addresses this} by adding missing motion using 2D trajectory inputs. 
In contrast, our method explicitly extracts and analyzes 3D information, enabling accurate shape editing, object insertion, and replacement even when the input videos have significant view transformations. 

\paragraph{Sketch-Based Content Editing}

Sketch-based interfaces have been widely used in image generation \cite{Pix2Pix, SketchCoCo, CycleGAN, DeepFaceDrawing, T2I-Adapter, ControlNet, semantic_diffusion, DualSmoke}, video generation \cite{VidSketch, Cartoon_inter_TVCG, ToonCrafter, LVCD}, and 3D generation \cite{DeepSketch2Face, Sketch2Pose, Sketch_Depth, Sketch_SDF, Sketch3D, sketch_texture, survey_3DGen}. 
Compared to generation, editing requires preserving original features while producing reasonable modifications.
This challenge can be addressed as a sketch-guided inpainting problem in image editing \cite{DeepFill_v2, SketchEdit, Faceshop, SC-FEGAN, MagicQuill}.
Similar ideas have been utilized for video editing. 
{For example,} VIRES \cite{VIRES} designs a sequence ControlNet to enable sketch-based video edits but requires sketches and masks of edited regions for all frames.
FramePainter \cite{FramePainter} further utilizes video models for sketch/dragging-based image editing.
Instead of fast inference, sketch-based 3D editing methods \cite{SketchDream, SKED, SketchGS} rely on optimization techniques to modify 3D representations but only focus on objects instead of 3D scenes. 
Our method designs a 3D-aware approach to generate masks for each frame, {using a} 3D point cloud representation to handle videos with significant view changes. 
Additionally, our method generates realistic results across diverse scenarios, including both indoor and outdoor scenes, addressing the limitations of previous sketch-based editing techniques.

\section{Methodology}

\begin{figure*}[h]
    \centering
    \includegraphics[width=1.0\linewidth]{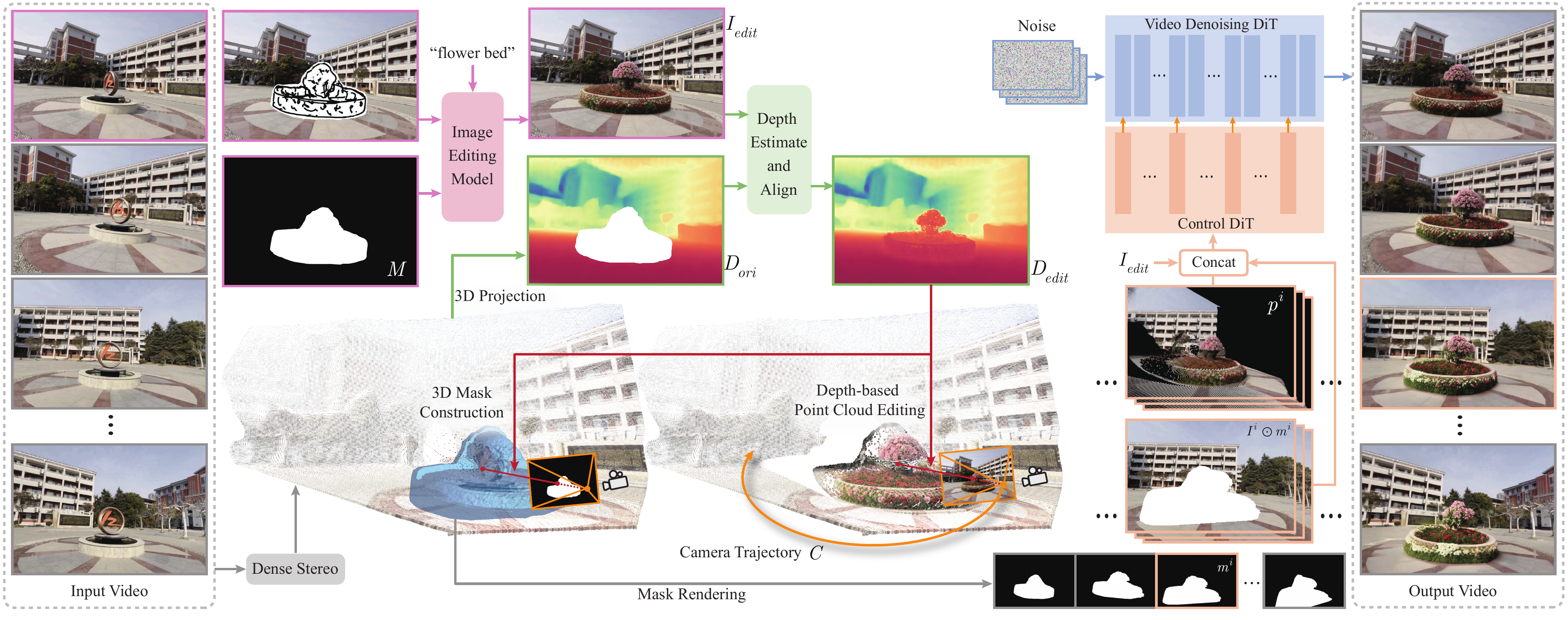}
    \caption{The pipeline of our sketch-based 3D-ware video editing method. 
    Given an input video, the edited result of the first frame, %
    $I_{edit}$ is generated by the image editing model \cite{MagicQuill} using the sketch, mask, and text prompt input. 
    We derive a 3D point cloud and camera parameters from the input video to propagate the editing. Then, we utilize a depth map to represent and align the newly edited content with the original scene to generate the edited point cloud. 
    To identify edited regions, we construct a 3D mask that is rendered for all the frames to generate a mask sequence. 
    The point cloud rendering results, the original video containing unedited regions, and the edited images serve as inputs for the conditional video diffusion model to synthesize the final edited video. Input Video ©DL3DV-10K.
    }
    \Description{pipeline}
    \label{fig:pipeline}
\end{figure*}

Given an input video, users incorporate hand-drawn sketches, masks, and text prompts to indicate desired modification effects. 
Our objective is to ensure the newly generated content remains consistent with the original video's view changes while accurately preserving unedited regions. 
We assume the video captures static primarily scenes, mainly with significant viewpoint changes. 
During editing, we leverage MagicQuill \cite{MagicQuill}, a sketch-based image editing method (detailed in Sec. \ref{sec:preliminary}), to edit the first frame. 
As shown in Fig. \ref{fig:pipeline}, to propagate these edits across the video, our method extracts 3D information from the input video, including a point cloud and camera poses.
We introduce a point cloud editing approach, which utilizes depth maps to represent and align the 3D geometry of the edited regions, as discussed in Sec. \ref{sec:point_cloud_editing}. 
Additionally, Sec. \ref{sec:mask_propagation} presents our video mask propagation strategy, designed to track the edited regions across all frames.
Finally, in Sec. \ref{sec:diffusion_video}, we develop a video diffusion model that synthesizes realistic edited videos with input conditions of point cloud renderings, the original video, and propagated masks.
This pipeline ensures coherent and realistic editing across diverse view transformations, bridging 2D edits and 3D-aware video generation.

\subsection{Preliminary}\label{sec:preliminary}

\paragraph{Image Editing Model.}
{MagicQuill \cite{MagicQuill} generates realistic edited images from an input image, text prompt, mask, and a sketch with optional color strokes.
It builds on Stable Diffusion \cite{StableDiffusion} with two additional networks: ControlNet \cite{ControlNet} for effective editing and an inpainting U-Net \cite{BrushNet} for preserving unedited regions. 
The model is trained to reconstruct images under conditions that imitate the editing process. 

\paragraph{Video Diffusion Model.}
CogVideoX \cite{cogvideox}, a state-of-the-art text-to-video generation model, is utilized for video synthesis. 
It employs a 3D causal VAE for video compression into a latent space, followed by a denoising process to generate latent codes.
The DiT architecture, which contains multiple transformer blocks, is utilized to generate high-quality videos with good temporal consistency.
Our video generation model adds an additional control block based on the CogVideoX model.
}

\subsection{Depth-based Point Cloud Editing}\label{sec:point_cloud_editing}
Given a video with frames $I = \{I^1, I^2, ..., I^N\}$, where $N$ is the number of frames, the first frame is edited by MagicQuill to generate image $I_{edit}$.
Directly applying an image-to-video generation model on $I_{edit}$ cannot control camera movement, leading to totally different viewpoint changes from the input video. 
To address this, we argue that explicit 3D geometry analysis and manipulation is a promising solution.
Following \cite{ViewCrafter}, we use point clouds as {a 3D representation} %
to guide video generation, ensuring that the viewpoint changes of new content align with the original video.

We employ DUSt3R \cite{DUSt3R}, a dense stereo model, to %
obtain {a 3D point cloud and camera parameters} %
from the input frames. 
Given two images $I^1, I^2$ from different views, DUSt3R predicts point maps $O^{1,1},O^{2,1}$ in the same camera coordinate system of view $I^1$.
This process is repeated between all \hb{pairs of frames} %
in the video, followed by global point map alignment to obtain {\hb{the} output 3D information.} %
Please refer to \cite{DUSt3R} for more details. 
We denote the point cloud of the first frame as $P$ and the camera parameters for all frames as $C = \{C^1, C^2, ..., C^N\}$.

To effectively propagate the edits in $I_{edit}$ across novel view frames, we translate the edited image $I_{edit}$ into an edited point cloud $P_{edit}$, which is rendered with camera $C$ to provide the pixel level guidance for novel view frame generation.
To achieve this, we find that depth maps effectively encode the relative {geometry} relationship {among objects} within {a} 3D scene and explicitly have correspondence {between the pixels in unedited regions}, which can be used for point cloud alignment and editing.

Specifically, with the point cloud $P$ and camera parameters $C^1$, the depth map of the first frame $D_{ori}$ before editing is generated by 3D projection. 
Similarly, the depth map of $I_{edit}$ can also be obtained with a new point cloud and camera estimated by DUSt3R (with $I_{edit}$ duplicated to imitate paired input, as in \cite{ViewCrafter}). Depth estimation model\hb{s like} \cite{DepthAnything_v2}
can also be optionally used to get {depth maps}.
We normalize the values of the edited depth map to get $\hat{D}_{edit}$. 
Then, two alignment coefficients, translation $t$ and scale $s${, are} required to transform the relative depth maps $\hat{D}_{edit}$ into the absolute one like $D_{ori}$. 
Notably, the geometry in unedited regions $\overline{M}$ {before and after editing} naturally has %
correspondence, %
which can be used to obtain these coefficients.
We utilize a simple least squares algorithm to achieve it. 
The optimization objective is to minimize the pixel distances:
\begin{equation}
    E(s,t) =  {\textstyle \sum_{i=1}^{m}} \left \| (s\hat{d}_i+t)-d_i \right \| ^2, 
\end{equation}
where $d_i$ and $\hat{d}_i$ are depth values in unedited regions of $D_{ori}$ and $\hat{D}_{edit}$, respectively, and $m$ is the number of pixels in unedited regions $\overline{M}$. 
Users can also adjust %
$t,s$ to obtain the best performance.
We merge the edited regions into the original depth maps to get the final edited depth maps:
\begin{equation}
    D_{edit} = D_{ori} \odot \overline{M} + (\hat{D}_{edit} \cdot s + t) \odot M.
\end{equation}
The final edited point cloud $P_{edit}$ is obtained by back-projection:
\begin{align}
P_{edit}=o^1 + r^1 \cdot D_{edit}, 
\end{align}
where $o^1$ is the ray origin and $r^1$ is the ray direction corresponding to camera $C^1$.

\subsection{Video Mask Propagation}\label{sec:mask_propagation}

For ease of interaction, we require users to draw a 2D mask $M$ only in the first frame to label edited regions. This requires us 
to track the mask in novel view frames to merge new edited content with the original video while accurately preserving unedited regions.
This is nontrivial because the mask shapes are dynamically changed for each frame and must %
{align} with {the original} view transformation.

{We address this by leveraging the depth maps before and after editing as priors.
Specifically, we utilize a mesh model to build a 3D mask. 
We start with a cylindrical shape, whose top and bottom surfaces are defined by back-projecting 2D mask $M$ along camera rays. 
This ensures the rendered 3D mask matches the 2D mask's shape in the edited view. 
To refine the 3D mask's geometry, for the top surface, we merge pre- and post-edit geometries by minimizing depth values $D_{ori}$ and $D_{edit}$ for each pixel.
Then, vertex positions are adjusted based on this merged depth map, while edges are constructed according to pixel adjacency.}
For simplicity, we assume the %
{bottom surface remains} %
a plane structure, constructed with the uniform {depth values obtained by the maximum of} %
the edited regions in $D_{ori}$ and $D_{edit}$. 
A side face further connects the frontal and back {surfaces.}
With the camera $C$, we render the 3D mask to get a sequence of 2D masks, %
$m^1, m^2, ...., m^N$, which are utilized for precise region modification in video editing.
More details of the 3D mask construction are discussed in the supplemental material.

\subsection{Diffusion-based Video Generation}\label{sec:diffusion_video}

A video diffusion model generates the final realistic edited videos. 
We render the edited point cloud $P_{edit}$ with camera parameters $C$ to obtain conditions $p=\{p^1, p^2, ..., p^n\}$.
The point cloud rendering results encode the camera transformation and enable explicit propagation of the editing manipulation. %
The input video (with the edited regions %
masked) also serves as {a condition} %
to preserve the original features. 
Additionally, we add the edited image $I_{edit}$ as {a} condition for each frame to provide {the} appearance {reference and thus} %
enhance the video details. 
We utilize a modified ControlNet \cite{ControlNet} branch to insert the camera control and original video information.
The input of the control branch {is formulated} as,
\begin{align}
    c_{edit} = \{c_{edit}^i\}_{i=1}^{N} = \{concat(p^i,I^i \odot m^i, I_{edit})\}_{i=1}^{N}.
    \label{eq:edit_condition}
\end{align}
This video diffusion model removes the artifacts of point cloud rendering by inpainting the missing holes in rendering results, and corrects the slight geometry distortion caused by point cloud estimation. 
It also analyzes the relationship between the newly generated components and the original video to achieve seamless fusion. 

\paragraph{Training Strategy.}
{We use} a self-supervised training strategy %
to train the video diffusion models. 
We create paired data, including %
input frames $I$, the first frame's point cloud rendering {images} $p$, %
and randomly generated 2D mask $M$ {imitating user-drawn cases.} %
We utilize the mask propagation approach in Sec. \ref{sec:mask_propagation} to obtain a mask sequence $\{m_i\}_{i=1}^{N}$. 
These conditions are concatenated as the input condition signals $c$ similar to Eq. \ref{eq:edit_condition}, with the first frame $I^1$ as the editing input of $I_{edit}$. 
The proposed training strategy is the same as {that used for} video inpainting but with explicit 3D guidance. 
The video denoising transformer network $\epsilon _{\theta }$ is optimized by the following diffusion loss:
\begin{align}
    L(\theta )=\mathbb{E} _{t\sim u(0,1),\epsilon \sim N(0,1)}[\left \| \epsilon _{\theta } (z_t,t,y,c) - \epsilon \right \| _{2}^{2} ], 
\end{align}
where $z_t=\alpha_t z_0 + \sigma_t \epsilon$, with $z_0$ being the encoded ground-truth latent code of frames $I$. $\alpha_t$ and $\sigma_t$ are hyperparameters of the diffusion process.
The model takes an additional input of text prompt $y$ detected by LLaVA \cite{LLaVA} during training and inference. 

\begin{figure*}[h]
    \centering
    \includegraphics[width=1.0\linewidth]{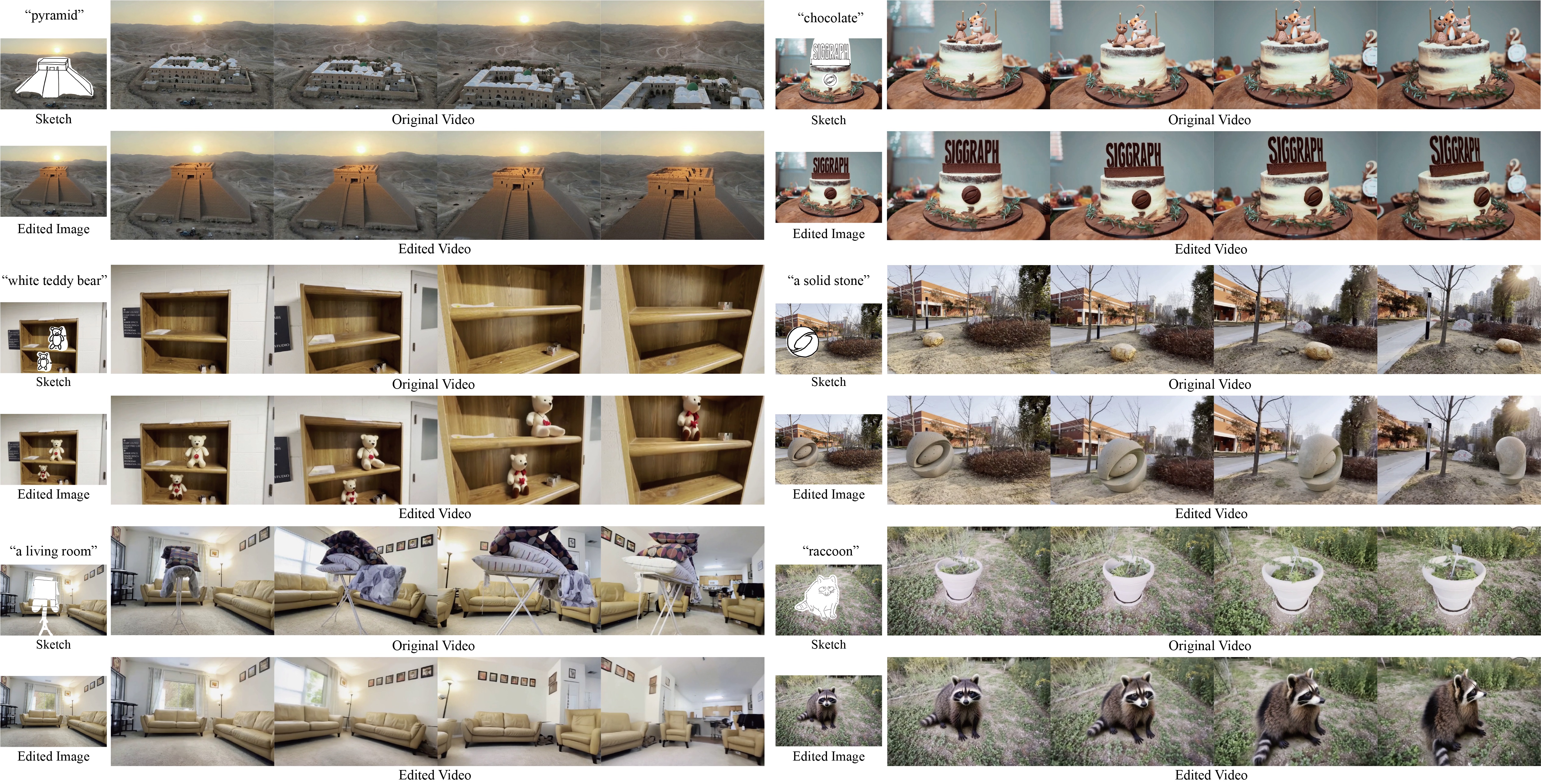}
    \caption{Sketch-based video editing results. 
    For each example, we show the {input text prompt, sketch, and edited image} on the left and the original and edited videos on the right. %
    Our method generates realistic editing results for videos with significant viewpoint changes. Original Video ©DL3DV-10K, ©Taryn Elliott, ©Samir Smier.
    }
    \Description{results}
    \label{fig:results}
\end{figure*}

\section{Evaluation}

In Sec. \ref{sec:results}, we show the video editing results of our method.
We have conducted extensive qualitative and quantitative experiments, including comparison (Sec. \ref{sec:comparison}), ablation study (Sec.\ref{sec:ablation_study}), and user study (Sec. \ref{sec:user_study}) to validate the superiority of our method compared with existing or baseline approaches. 

\paragraph{Implementation Details.}
Our video diffusion model is built based on CogVideoX-2b \cite{cogvideox}, with an additional conditional network \cite{ControlNet} composed of 10 control blocks. %
We fixed the weights of base video generation model while updating the condition network.
Our model is trained on the DL3DV \cite{DL3DV} and RealEstate \cite{RealEstate} datasets. 
The training pairs include videos, point cloud renderings, and random masks. 
We first train the control network with only point cloud rendering inputs with 20,000 steps and then with additional masked video inputs for an additional 15,000 steps.  
More details on the network, training, and data processing %
are in the supplemental materials.

\subsection{Results}\label{sec:results}
Our method enables detailed {sketch-based editing} for {scene} videos with significant viewpoint changes. 
As shown in Fig. \ref{fig:results}, {given user-drawn} sketches, masks, and text prompts, our method {automatically} synthesizes realistic {edited} videos. %
The video diffusion model generates high-quality novel view results for editing regions while accurately preserving the unedited regions. 
Thanks to our 3D analysis and condition guidance, the newly generated parts exhibit reasonable view transformations and seamlessly merge with the original scene. 
Our method supports diverse editing operations, including object insertion, replacement, removal, and component modification. 
{It works for} input videos with {various} camera transformations, including rotation, translation, and zoom variation. %
Please refer to the supplemental video {to examine edited video results}. %

\subsection{Comparison}\label{sec:comparison}

To validate the superiority of our method, we compare it with existing %
techniques. Specifically, 
we compare with video editing methods, including AnyV2V \cite{AnyV2V} and I2VEdit \cite{I2VEdit}, which propagate the editing of the first frame to the entire video.
We also compare with camera-controllable video generation methods, including ViewExtra \cite{ViewExtrapolation} and ViewCrafter \cite{ViewCrafter}, which take point cloud rendering as inputs (same as our method) and generate video results. 

\begin{figure*}[h]
    \centering
    \includegraphics[width=1.0\linewidth]{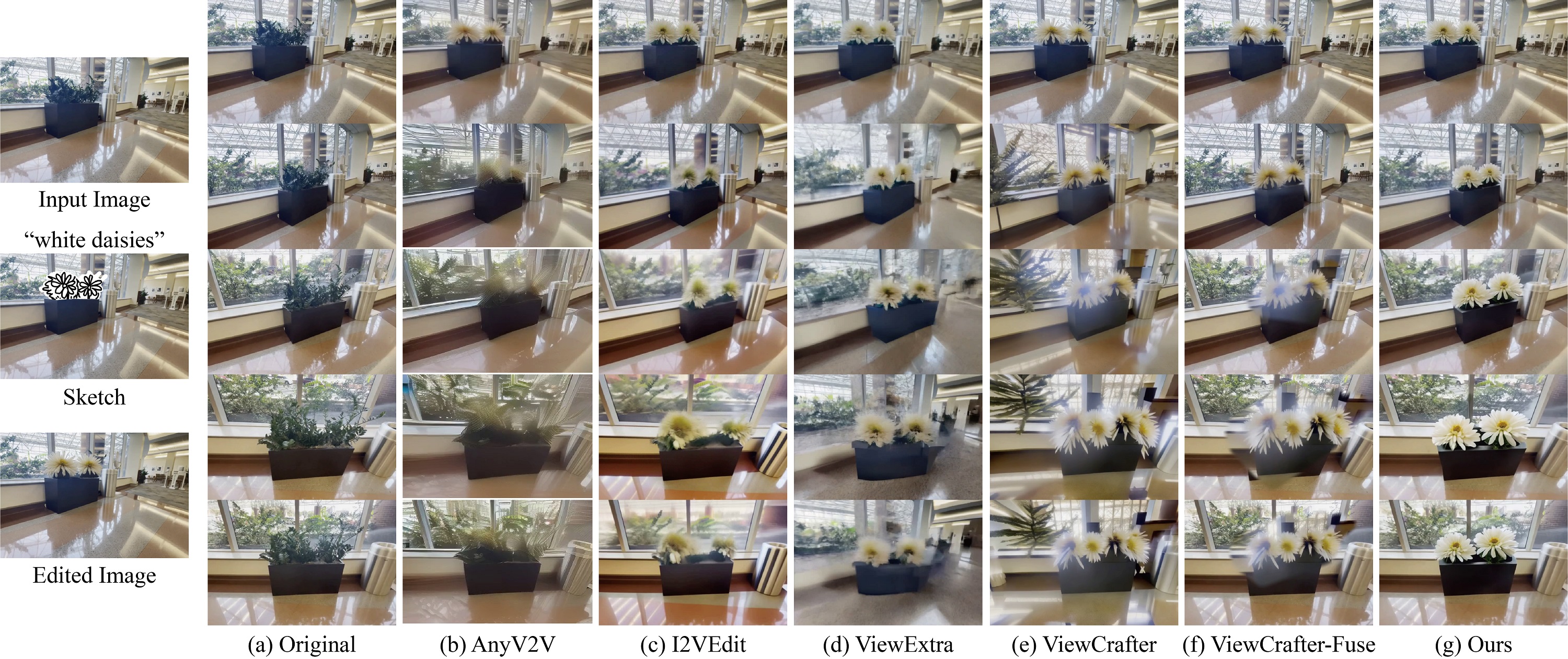}
    \caption{Comparison with existing approaches. Given %
    original video (a) and {sketch-based editing on the first frame (Left)}, %
    {existing} video editing methods, including AnyV2V \cite{AnyV2V} and I2VEdit \cite{I2VEdit}, generate fuzzy details. 
    3D-aware video generation methods, including ViewExtrapolation \cite{ViewExtrapolation} and ViewCrafter \cite{ViewCrafter} cannot handle significant viewpoint changes. 
    Directly fusing the generation results of Viewcrafter with the original video in pixel space {improves the results but with} obvious seams (f). In contrast, 
    our method (g) generates the most realistic video editing results. Original Video ©DL3DV-10K.
    }
    \Description{compare}
    \label{fig:compare}
\end{figure*}

AnyV2V \cite{AnyV2V} is a training-free method that merges the features of the original video with edited images. It is effective for appearance editing, but {encounters} information conflict for geometry editing, resulting in shape blending of leaves and flowers in Fig. \ref{fig:compare} (b).
I2VEdit \cite{I2VEdit} utilizes trainable LoRa to %
distill motion from input video, generating results (Fig. \ref{fig:compare} (c)) with clear shapes %
{but exhibiting} fuzzy details in frames distant from the edited image. 
ViewExtrapolator \cite{ViewExtrapolation} (shortened as ViewExtra) is a training-free method that refines the point cloud rendering with diffusion prior but struggles with large viewpoint changes, leading to frame quality degradation, as shown in Fig. \ref{fig:compare} (d). 
ViewCrafter \cite{ViewCrafter} fine-tunes the video diffusion model to accept additional input of point cloud rendering, making it more robust for view changes. 
However, the generated components (e) exhibit strange geometry, and the unedited regions are completely changed. 
Merging the results into the input video (f) preserves the unedited regions, but the editing regions are inconsistent and display obvious seam artifacts.
Our method (g) generates high-quality video editing results with clear details and accurate preservation of unedited regions. 

\begin{table}[!t]\scriptsize
    \caption{The quantitative comparisons with existing methods. Our method outperforms existing approaches in all metrics.
    }
    \label{tab:editing_compare}
    \begin{center}
    \resizebox{\columnwidth}{!}{
    \begin{tabular}{lcc|cccc}
        \toprule
         & \multicolumn{2}{c}{\textit{Auto Evaluations}} & \multicolumn{4}{c}{\textit{Human Preference}}\\
        \textit{Method} & {CLIP$\uparrow$} & {PSNR$\uparrow$} & VC & UP & TC & EQ \\\midrule
        AnyV2V \quad
        & 92.28 & 20.09 & 4.21 & 4.18 & 4.25 & 4.40 \\
        I2VEdit \quad
        & 94.92 & 22.31 & 3.23 & 3.05 & 3.18 & 3.25 \\
        ViewExtra \quad
        & 94.57 & 17.76 & 3.75 & 3.83 & 3.75 & 3.74 \\
        ViewCrafter \quad
        & 94.95 & 17.94 & 2.68 & 2.80 & 2.66 & 2.48 \\
        Ours & \textbf{95.47} & \textbf{31.20} & \textbf{1.14} & \textbf{1.15} & \textbf{1.15} & \textbf{1.12} \\
    \end{tabular}}
    \end{center}
\end{table}

\paragraph{Quantitative Comparison.}
We further employ automatic metrics to measure the performance of the compared approaches. 
Following %
\cite{ReVideo}, we utilize the CLIP \cite{CLIP} score between two consecutive frames to assess temporal consistency. 
The PSNR metric in unedited regions is used to measure the preservation of input videos. 
We randomly select %
20 examples from all our editing cases to calculate the above-mentioned metrics. 
As shown in Table \ref{tab:editing_compare}, our method significantly outperforms other methods in PSNR, validating that our method effectively preserves the features of unedited regions. 
Our method also outperforms existing methods in CLIP similarity, supporting the best temporal consistency.

\subsection{Ablation Study}\label{sec:ablation_study}
We conducted {an ablation study} %
to validate the necessity of our key components. 
In our method, depth maps are utilized to represent the geometry of edited regions, followed {by} %
scale and shift transformations to align with the original 3D scene. 
Since we utilize DUSt3R \cite{DUSt3R} to estimate the point cloud of the edited images, an intuitive baseline is to directly use traditional alignment approaches, such as ICP \cite{ICP}, to align the new point cloud with the original point cloud in 3D space. 
During editing, since only the unedited regions have correspondences, we calculate the rotation and translation based on the edited point cloud in unedited regions and then apply transformations to the whole edited point cloud. 
However, for this baseline, the edited point cloud has a slightly different scale from the original one, and the correspondences of unedited regions are not considered.  
As shown in Fig. \ref{fig:ablation1} (b), without the depth-guided alignment, the point cloud rendering results have holes in the first frame, and the new edited content is in the wrong %
position, resulting in a strange geometry of the new content. 
Our method generates reasonable point cloud rendering (d) with realistic video editing results (e).

\begin{table}[!t]\scriptsize
    \caption{The quantitative results of the ablation study. Our method outperforms all baseline approaches. {LPIPS values are scaled by a factor of 100.}
    }
    \label{tab:editing_ablation}
    \begin{center}
    \resizebox{\columnwidth}{!}{
    \begin{tabular}{ccccc}
        \toprule
        \textit{Metric} & w/o depth & w/o mask & w/o point cloud & ours \\
        \midrule
        PSNR $\uparrow$ \quad
        & 25.87 & 18.29 & 23.62 & \textbf{26.88}  \\
        LPIPS $\downarrow$ \quad
        & 9.12 & 20.35 & 9.65 & \textbf{8.28}  \\
    \end{tabular}}
    \end{center}
\end{table}

\begin{figure*}[h]
    \centering
    \includegraphics[width=1.0\linewidth]{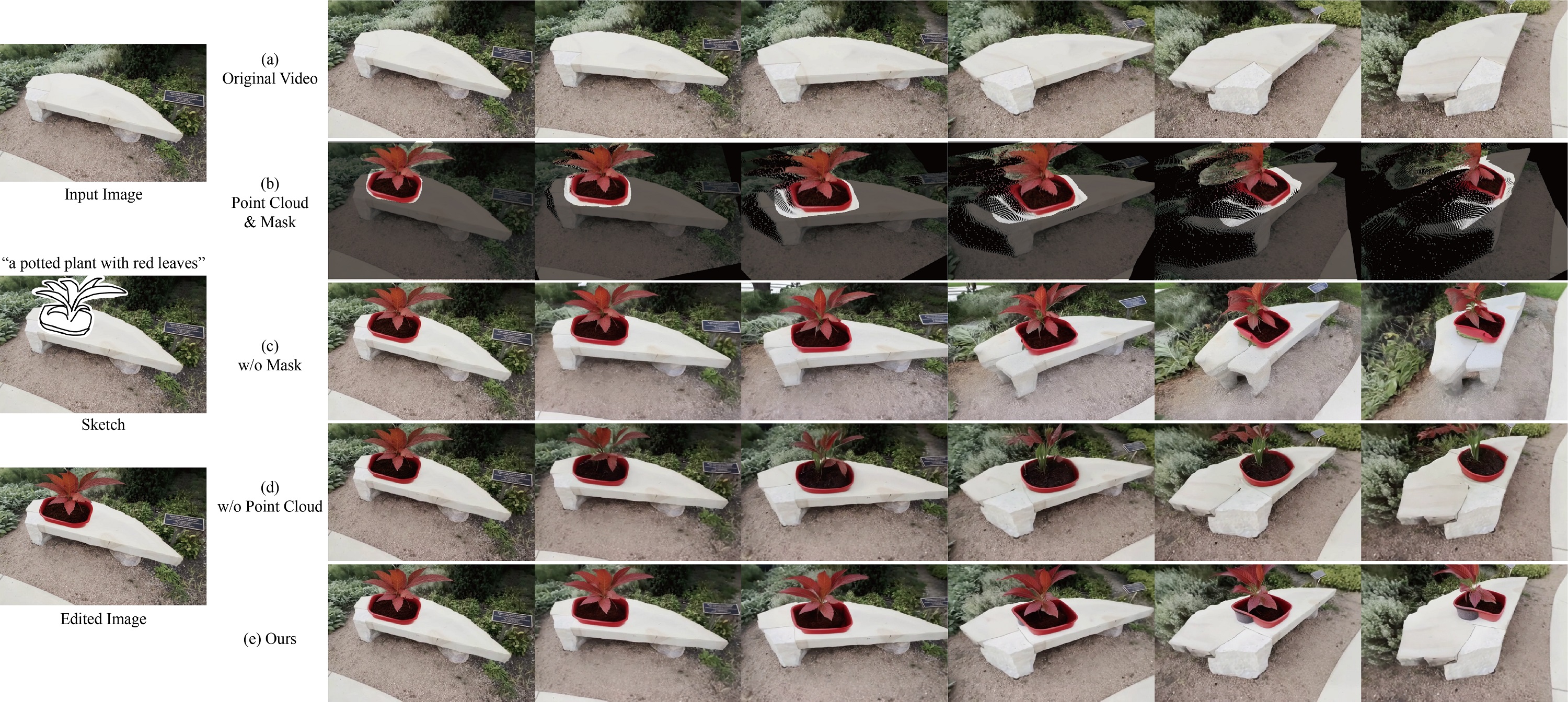}
    \caption{The ablation study of 3D-aware mask and point cloud rendering condition. 
    Given the original video (a) and editing operations (Left), our method produces the edited point cloud rendering and 3D-aware masks (b). 
    Without the masks (c), the unedited regions \lfl{(e.g., the shape of stone)} are not preserved. 
    Without the point cloud rendering (d), the shape of red leaves is changed in different frames. 
    Our method (e) generates the best results. Original Video ©DL3DV-10K.
    }
    \Description{ablation2}
    \label{fig:ablation2}
\end{figure*}

During video generation, the masked original video and point cloud rendering serve as conditions to control the unedited and edited regions, respectively. 
As shown in Fig. \ref{fig:ablation2}, if we remove the conditions of 3D-aware masks and the original video, the results (c) cannot preserve the original features, such as the different shapes of stones and green leaves, because these occlusion components are imagined by the models. 
If we remove the point cloud rendering (d), the explicit 3D guidance is missing, and thus, the complicated structure (shape of red leaves) cannot be maintained during viewpoint changes. 
Our method (e) generates view-consistent new content and accurately preserves the unedited regions. 

We further conduct quantitative experiments to validate the effectiveness of each component. 
We randomly select 30 test examples from DL3DV and RealEState and estimate the point cloud and camera parameters.
We detect object masks using SAM \cite{segmenet_anything} and then remove a random object in the point cloud and original video to imitate the editing process.
We treat the first frame of the original videos as edited images and then utilize our method to reconstruct the original videos. 
As shown in Table \ref{tab:editing_ablation}, our method achieves %
the best reconstruction values, validating %
that our method outperforms the %
baseline approaches.

\subsection{User Study}\label{sec:user_study}

We conducted a user study to visually compare our method with the existing approaches. 
We randomly selected %
10 editing examples and invited 24 people to participate in our questionnaires.
They ranked different methods in four aspects: viewpoint change consistency with the original video (VC), unedited preservation (UP), temporal consistency (TC), and overall editing quality (EQ).  
The rankings provided by the participants were used as scores.
As shown in Table \ref{tab:editing_compare}, our method outperforms others in all criteria, validating the superior performance of our method.

\section{Applications}

\subsection{Camera-Controllable Video Generation}\label{sec:3D_generation}

\begin{figure*}[h]
    \centering
    \includegraphics[width=0.95\linewidth]{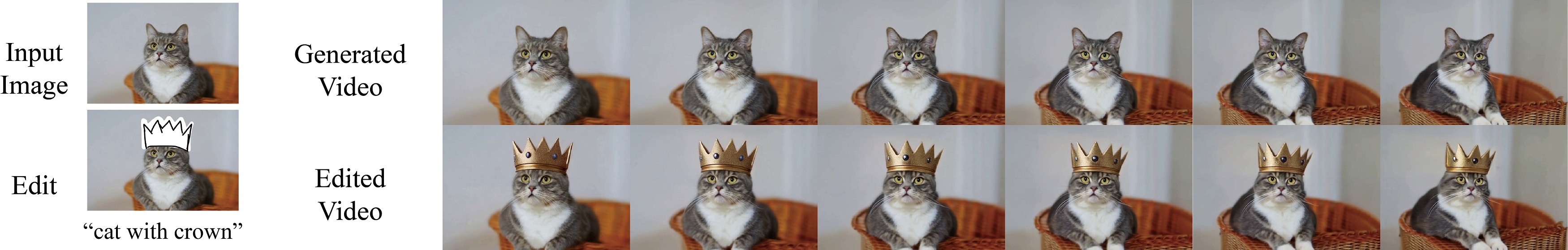}
    \caption{The application of camera-controllable video generation. 
    Given a single input image, users can control the camera viewpoints and generate a video with novel view results. Local region editing can be further added to achieve detailed, controllable video generation. Input Image ©Cats Coming.
    }
    \Description{application_3D_gen}
    \label{fig:application_3D_gen}
\end{figure*}

As shown in Fig. \ref{fig:application_3D_gen}, given a single image, similar to ViewCrafter \cite{ViewCrafter}, users can define a camera trajectory to produce point cloud rendering results, which serve as conditions %
to achieve camera-controllable video generation. 
Users can further edit the input image to update the generated videos while preserving unedited regions, showcasing detailed controllable generation ability.

\subsection{More Editing interaction tools}\label{sec:stroke_base_editing}

\begin{figure*}[h]
    \centering
    \includegraphics[width=0.95\linewidth]{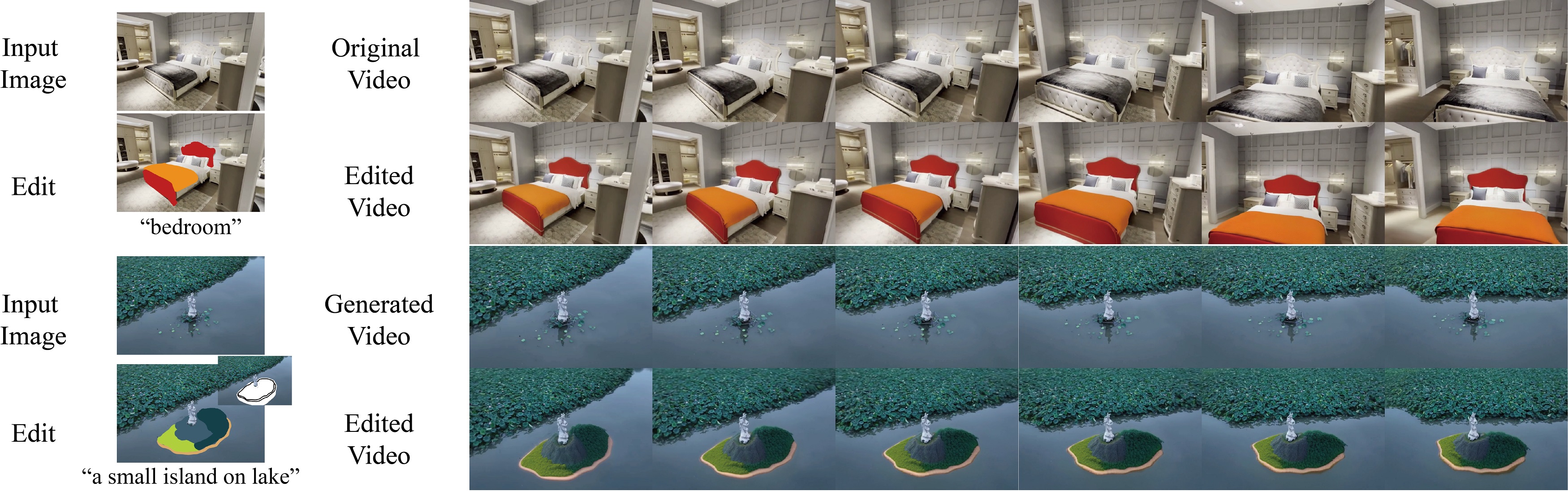}
    \caption{The application of color stroke-based video editing. 
    The appearance of existing content (bed in the 1st example) 
    can be edited by using color strokes. 
    Users can also control the appearance of newly generated objects, with the editing sketch shown in the top right corner in the 2nd example. Original Video ©DL3DV-10K.
    }
    \Description{application_color}
    \label{fig:application_color}
\end{figure*}

\paragraph{Stroke-based Video Editing.} As shown in Fig. \ref{fig:application_color}, users can draw color strokes to change the appearance of objects in images (supported by MagicQuill \cite{MagicQuill}), and the editing is reasonably propagated to the whole video. 
Additionally, as shown in the island insertion examples in Fig. \ref{fig:application_color}, the sketches and color strokes can be simultaneously input to control the geometry and appearance. %

\begin{figure*}[h]
    \centering
    \includegraphics[width=0.95\linewidth]{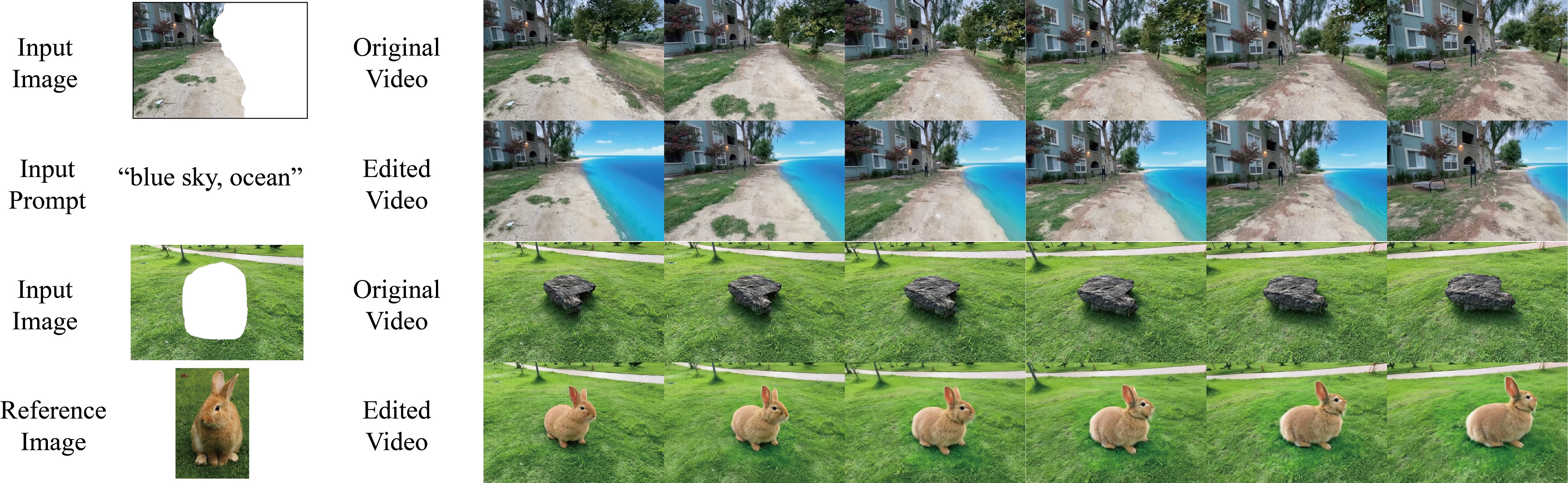}
    \caption{\rv{
    The application of image inpainting-based video editing. 
    These examples utilize the text-based \cite{FLUX_inpainting} and reference-based \cite{Paint_example} image inpainting to editing the first frame. The editing operations are well propagated into videos. Original Video ©DL3DV-10K, Reference image ©Pixabay.
    }
    }
    \Description{application_inpaint}
    \label{fig:application_inpaint}
\end{figure*}

\paragraph{Inpainting-based Video Editing}
Our method extends versatile image inpainting methods to 3D scene video editing. 
As shown in Fig. \ref{fig:application_inpaint}, text-based \cite{FLUX_inpainting} or reference-based \cite{Paint_example} image inpainting methods can be utilized to edit the first frame. 
These edits are then well propagated throughout the other frames to generate realistic video results.

\section{Conclusion}
In this paper, we have presented %
a novel sketch-based 3D-aware video editing method to change the %
{objects and components} %
in videos involving significant viewpoint changes. 
Our method supports diverse editing operations, including insertion, removal, replacement, and appearance modification of components and objects. 
The editing operations are applied to the first frame and propagated throughout the video. 
We utilize an explicit 3D point cloud representation to handle viewpoint changes and %
a depth-guided point cloud editing approach to represent the geometry of new content and align it with the original scene. 
To identify and preserve the unedited regions, we propose a 3D-aware mask propagation strategy, followed %
{by} a video diffusion model to generate the final realistic video editing results. 
Extensive experiments support the superior performance of our method compared with existing approaches. 

\begin{figure*}[h]
    \centering
    \includegraphics[width=1.0\linewidth]{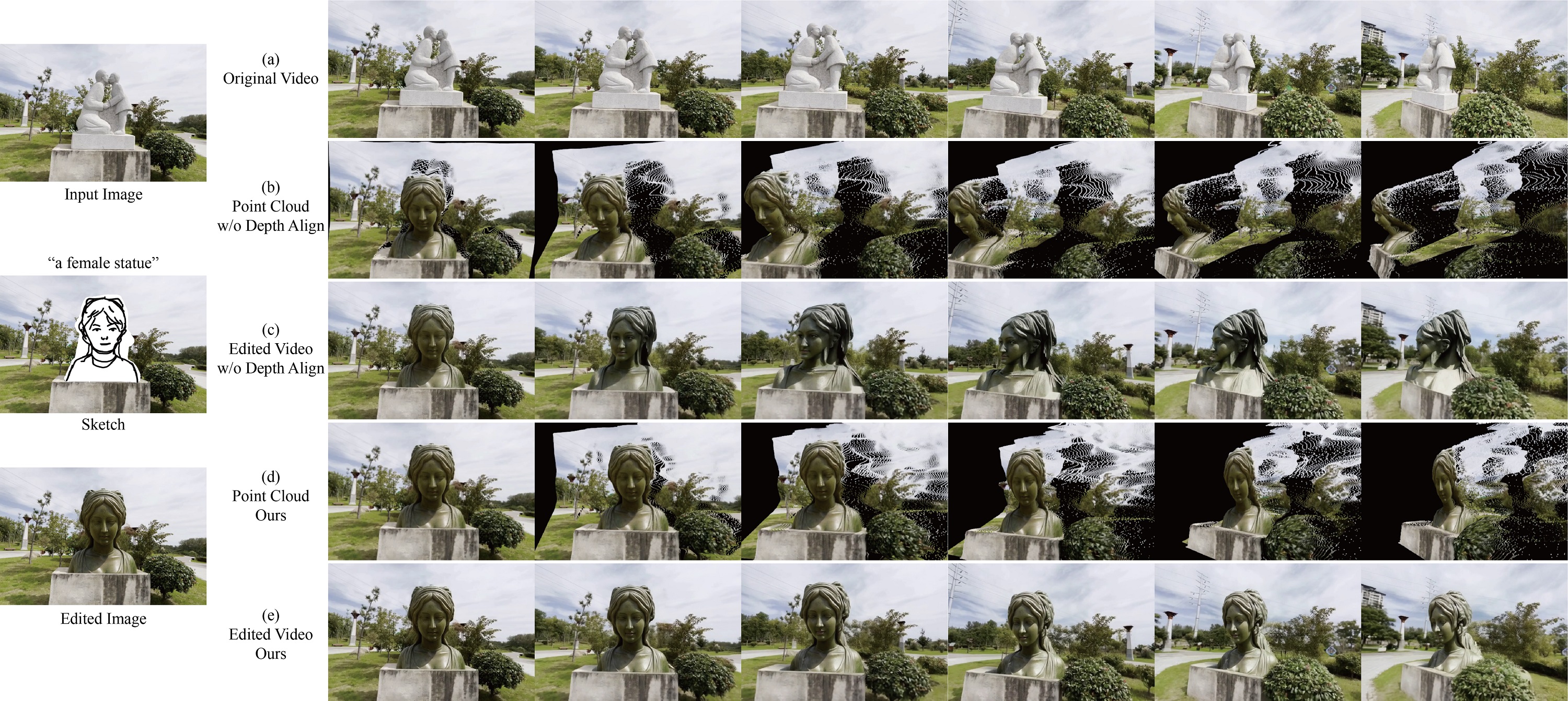}
    \caption{The ablation study of point cloud alignment. (a) An input video. (b, d) Edited point cloud rendering results. (c, e) Edited videos.
    Our method generates {a} more reasonable edited point cloud than {the} baseline without depth-guided alignment, resulting in better video editing results with reasonable geometry and %
    component interaction. Original Video ©DL3DV-10K.
    }
    \Description{ablation1}
    \label{fig:ablation1}
\end{figure*}

\begin{figure*}[h]
    \centering
    \includegraphics[width=0.92\linewidth]{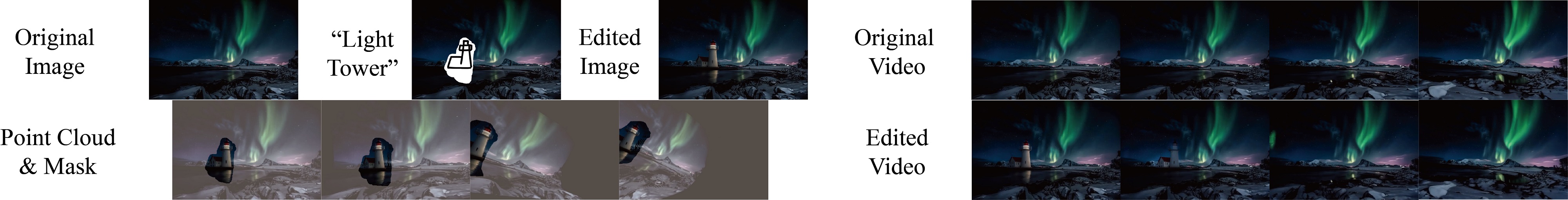}
    \caption{Failure cases. For the video that DUSt3R \cite{DUSt3R} estimates wrong point clouds and/or camera parameters, the edited point cloud and mask are also mistakenly rendered. The quality of the generated video is severely affected. Original Image ©stein egil liland. 
    }
    \Description{failure_case}
    \label{fig:failure_case}
\end{figure*}

\paragraph{Limitation and Discussion.}
Our method utilizes DUSt3R \cite{DUSt3R} to estimate the camera parameters and point cloud during 3D analysis. 
However, for challenging cases far from the training data of DUSt3R, as shown in Fig. \ref{fig:failure_case}, it predicts incorrect 3D information, which is inherited during the editing and causes obvious artifacts.
{More robust 3D analysis approaches \cite{MonST3R, MV_DUSt3R} or model fine-tuning with data augmentation might be adopted to address this issue.}
Additionally, due to the restriction of the training dataset, our method currently cannot handle cases with 360° viewpoint changes. Employing a large-scale dataset with progressive generation could possibly solve this problem. 
Moreover, one limitation of our method is the restriction on mostly static 3D scenes.
Processing videos of objects simultaneously with large dynamic motion and significant viewpoint changes is interesting research, but it is challenging because disentangling this information is required for reasonable propagation. 
Furthermore, our method might introduce slight blurriness in background details, such as the sofa in Fig. \ref{fig:results} and the trees in Fig. \ref{fig:ablation1}. 
A potential solution is to leverage more powerful video diffusion models.

\begin{acks}
This work was sponsored by Beijing Municipal Science and Technology Commission (No. Z231100005923031), Innovation Funding of ICT, CAS (No. E461020), and the National Natural Science Foundation of China (No. 62322210).
The authors would like to acknowledge the Nanjing Institute of InforSuperBahn OneAiNexus for providing the training and evaluation platform.

\end{acks}

\bibliographystyle{ACM-Reference-Format}
\bibliography{main}

\end{document}


\title{\sysName: Sketch-based 3D-Aware Scene Video Editing Supplemental Material}

\maketitle

\section{Overview}

In this supplemental material, we discuss details of mask propagation, add additional implementation details, validate the robustness for depth-based alignment, discuss the quantitative comparison with video generation methods, show more comparison results, show the results of shadows and reflections, discuss the limitation of camera motion, show results of TU-Berlin dataset, and discuss the robustness for size, position and sketch style.

\begin{figure}[h]
    \centering
    \includegraphics[width=1.0\linewidth]{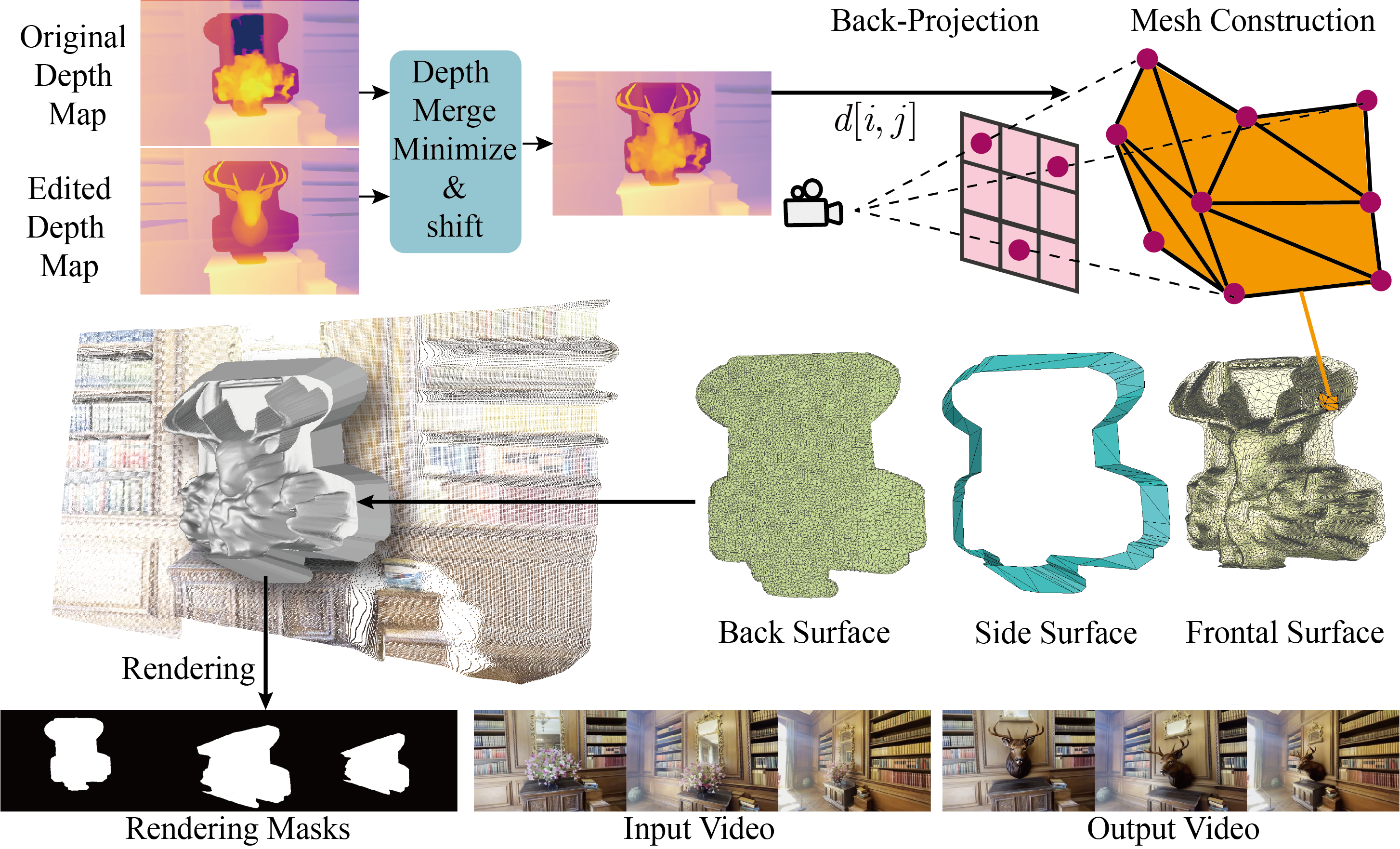}
    \caption{Detailed illustration of mask propagation. We utilize depth priors and 3D geometry information to achieve mask tracking. 
    Specially, the original and edited depth maps are merged by minimizing the pixel-wise values, followed by a slight shift to expand the 3D regions. 
    This merged depth map is then utilized to construct the frontal surface of the 3D masks by back-projection. The back surface is generated using a similar approach, with a uniform depth value. The side surface connects the frontal and back surfaces contours.
    These surfaces composes of a 3D mesh model, which is subsequently rendered as a video of the masks. Input Video ©DL3DV-10K.
    }
    \Description{3D_mask_supple}
    \label{fig:3D_mask_supple}
\end{figure}

\section{Details of mask propagation}
Given input videos, DUSt3R \cite{DUSt3R} estimates point clouds and camera parameters, enabling the extraction of the original depth map from the first frame.
With an editing sketch, text and 2D mask, the image editing and depth-based point cloud editing approach generate an edited depth map. 
These depth maps are then used as geometry priors for the 3D-guided mask propagation. 
As shown in Fig. \ref{fig:3D_mask_supple}, the original and edited depth maps are merged by minimizing the per-pixel values, forming a geometry union that including original and new contents.
The merged depth maps is further adjusted by subtracting a small value (default: 0.02) to dilate the mask geometry.

A mesh model, aligned with the point cloud in 3D space, is constructed to propagate the 2D masks into novel views. 
We utilize a cylindrical mesh template that consists of the frontal surface, back surface, and side surface. 
Specially, for each pixel $i,j$ within the 2D mask regions, we retrieve the corresponding depth value $d[i,j]$ along with the 3D camera ray's origin $o_{i,j}$ and direction $r_{i,j}$. 
This pixel is then back-project into the 3D space to obtain the vertex $o_{i,j} + d[i,j] * r_{i,j}$ in 3D space. 
All pixels within the 2D mask regions undergo this back-project to construct the vertices of frontal surface. 
Adjacent and diagonal pixels are then connect as edges to form triangle faces, as shown in the top right corner of Fig. \ref{fig:3D_mask_supple}.
A similar approach is utilized to construct the back surface, but with a uniform depth values due to the lack of detailed geometry in the back regions. 
The side surface connect the contours of the frontal and back surfaces.
Finally, the 3D mesh model is rendered in white color to obtain a sequence of masks $m^1, m^2,...,m^N$, which label the local edited regions in each frame.

\begin{figure*}[h]
    \centering
    \includegraphics[width=1.0\linewidth]{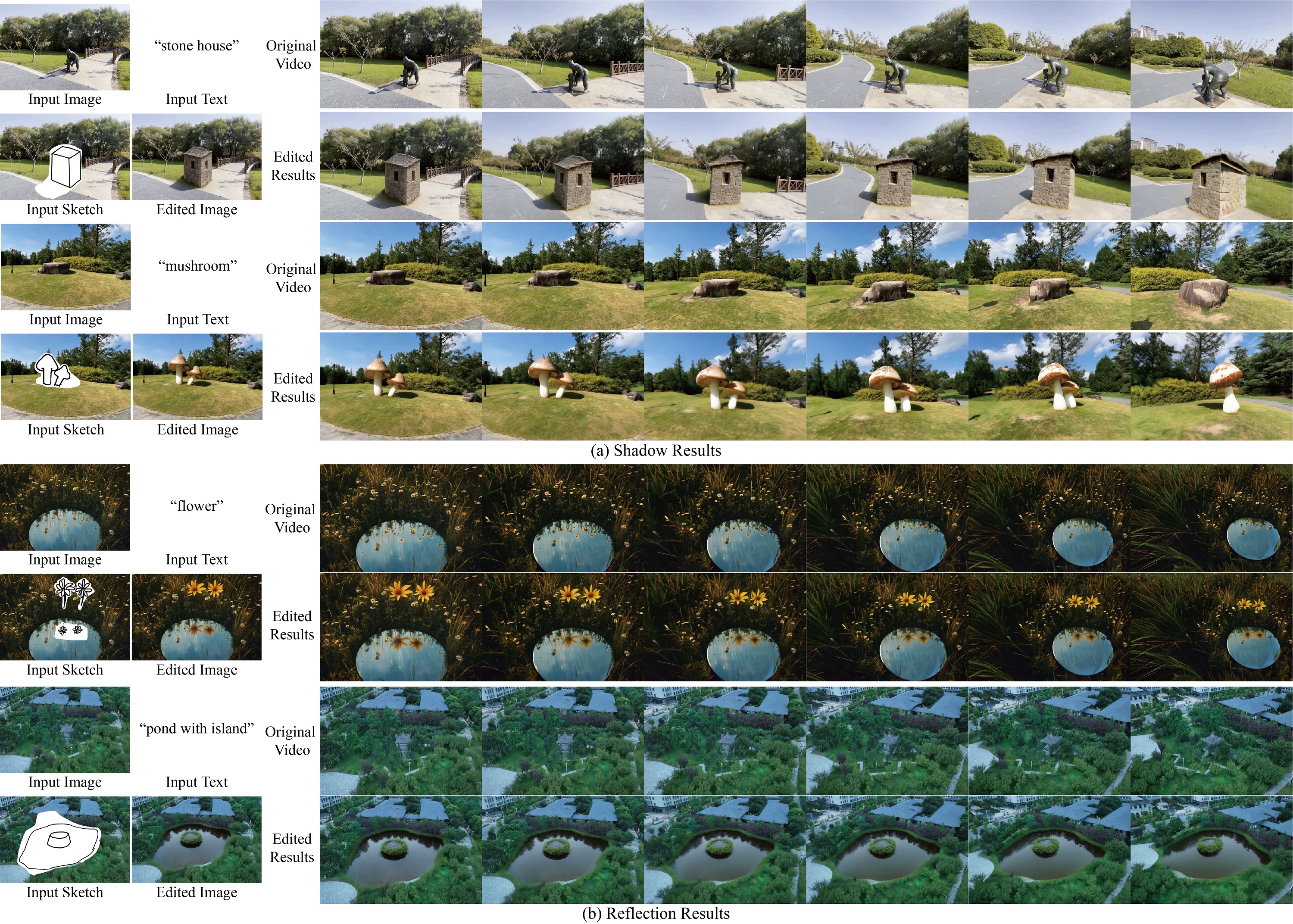}
    \caption{Results of editing objects with realistic shadows and reflections. The left side shows the input image, text, and sketches, while the right side displays the original and edited video frames. The edited images exhibit plausible shadow and reflection effects, which are consistently propagated across the video frames. Original Video ©DL3DV-10K, ©Lukáš Dlutko.
    }
    \Description{reflection_shadow}
    \label{fig:reflection_shadow}
\end{figure*}

\section{Additional Implementation Details}

We have enhanced CogVideoX-2B\cite{cogvideox} by adding an additional control network to incorporate 3D guidance and preserve unedited region from the original video, with similar network structure as \cite{ControlNet}. 
Initially, input conditions are encoded into the latent space by the pretrained VAE. 
These latent variables are then concatenated and processed via a feedforward layer and linear mapping to obtain hidden features, with the initial weights of linear mapping set to zero for fast convergence. 
The resultant hidden features are then feed into a DiT-based network comprising 10 blocks, which inject residual features at specific points (blocks 1, 4,..., 28) within the base model's intermediate hidden features.
Each block has the same structure and initial weight of its counterpart block in the base model, but includes an additional output linear mapping layer also initialized with zero weights. 

During the training, the weights of the base video generation model are fixed, while those of the control network are updated.
We utilize the AdamW optimizer with a learning rate of 5e-6. 
The control network is trained on three NVIDIA H800 GPUs, with a batch size of 1 and gradient accumulation of 4 for each GPU. 

\begin{table}[!t]
    \caption{
    Robustness of our video diffusion model for depth map alignment. 
    We use point clouds generated from various depth maps, including ground truth, noisy, and uniformly shifted versions in edited regions.
    Despite distortions in the depth maps, the synthesized videos maintain acceptable reconstruction quality. 
    LPIPS values are scaled by a factor of 100.
    }
    \label{tab:distortion_metrics}
    \begin{center}
    {
    \begin{tabular}{lccccccc}
        \toprule
        \textit{Metric} & GT Depth & Noise (20\%) & Shift (+20\%)  & Shift (-20\%) \\
        \midrule
        PSNR$\uparrow$ & 28.24 & 26.78 & 26.65 & 25.66 \\
        LPIPS$\downarrow$ & 7.3 & 9.5 & 9.5 & 9.9   \\
        \bottomrule
    \end{tabular}}
    \end{center}
\end{table}

\begin{table}[!t]
    \caption{
    Comparison with point cloud conditions video generation models, including ViewExtrapolation\cite{ViewExtrapolation} and ViewCrafter\cite{ViewCrafter}. 
    Our method has the best reconstruction quality, supporting the superior video generation of our model. LPIPS values are scaled by a factor of 100.
    }
    \label{tab:distortion_metrics}
    \begin{center}
    {
    \begin{tabular}{lccc}
        \toprule
        \textit{Metric} & ViewExtrapolation & ViewCrafter & Ours  \\
        \midrule
        PSNR$\uparrow$ & 17.94 & 18.27 & \textbf{19.09}  \\
        LPIPS$\downarrow$ & 28.95 & 21.61 & \textbf{16.93}   \\
        \bottomrule
    \end{tabular}}
    \end{center}
\end{table}

Our approach outputs videos with a resolution of 720×480, matching the outputs of CogVideoX-2B\cite{cogvideox}.
To facilitate point cloud and camera estimation, we resize the video into 512×336 before being processed by DUSt3R\cite{DUSt3R}. 
The edited point cloud is rendered back the resolution of 512×336 using the detected camera intrinsic and extrinsic parameters, and then resized into 720×480 as input conditions for the video diffusion model.

Our work leverages RealEstate\cite{RealEstate} and DL3DV\cite{DL3DV} datasets, which are preprocessed into a training dataset containing 111,332 six-second video clips with 49 frames. 
To construct paired dataset, we utilize DUSt3R\cite{DUSt3R} to 
estimate the point cloud of the first frame, and camera parameters for selected interval frames (1, 4, 8,..., 49), due to the constraint of computation time and memory. 
Intermediate frames camera parameters are obtained by interpolation. 
Two strategies are applied for generating random masks that mimic user editing: one involves detecting object masks using SAM\cite{segmenet_anything} followed by random selection and dilation, while the other adds random stroke trajectories onto images to create mask images.

\begin{figure*}[h]
    \centering
    \includegraphics[width=1.0\linewidth]{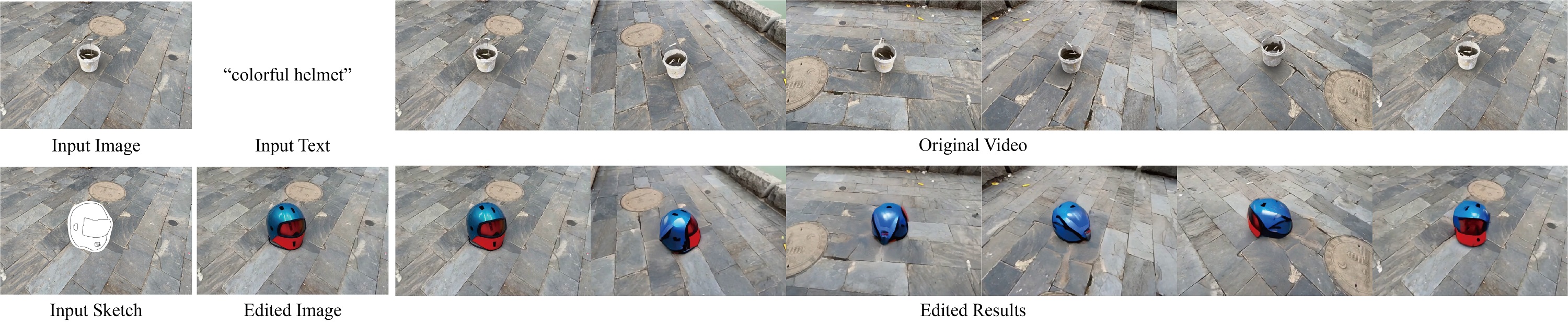}
    \caption{Unsuccessful results for camera motion involving 360-degree rotations. The left side shows the editing operations, while the right side presents the editing results. When the camera rotate back into the view of the edited frame, the results do not maintain original features, such as the distorted eyeglasses. Original Video ©DL3DV-10K.
    }
    \Description{failure_case}
    \label{fig:failure_case}
\end{figure*}

\section{Robustness for Depth-based Alignment}

We conducted an experiment to validate the robustness of our video diffusion model for depth-based point cloud alignment. 
Given examples in our ablation study, we obtain ground truth depth maps of the first frame using point cloud estimation and 3D projection. 
The selected object regions in the first frames (the same as our ablation study) are treated as editing regions. 
We determine the scale value of depth maps in these regions (maximum value minus minimum value), and then introduce distortions to the depth maps using different strategies:
adding uniform noise with a maximum value of 20\% scale units, and shifting the depth maps forward or backward with distance of 20\% scale units. 
Point clouds are reconstructed from these distorted depth maps, and rendered as input conditions for our video diffusion models.
We then calculate reconstruction metrics (PSNR and LPIPS) between the generated and ground truth videos. 
As shown in Table. \ref{tab:distortion_metrics}, even with distorted depth maps, the generated videos maintain acceptable reconstruction quality (PSNR>25).
This demonstrates the robustness of our video diffusion model for depth-based point cloud alignment. 

\section{Video Generation Comparison}

Our method supports the camera controllable video generation by using point cloud rendering as conditions. 
We compare our method with two similar video generation methods, ViewExtraplolation\cite{ViewExtrapolation} and ViewCrafter\cite{ViewCrafter}, which also use point cloud rendering as input. 
We randomly selected 20 examples from the test set of RealEstate\cite{RealEstate} dataset, then used the first frame's point cloud rendering as conditions to reconstruct these videos. 
The DL3DV dataset was not used due to the lack of a corresponding test set for ViewCrafter.
As shown in Table. \ref{tab:distortion_metrics}, our method achieves the best reconstruction performance, indicating that our generated videos are closest to real videos.

\section{Results of shadows and reflections}
Our method generates edited examples with realistic shadows and reflections effects.
As shown in Fig. \ref{fig:reflection_shadow} (a), the input videos contain strong sunlight and well-defined shadows.
After modifying the sketches, the newly added objects produce plausible shadows that blend seamlessly with the surroundings in edited images, including the self-occlusion shadows on mushrooms.
When the editing operations are propagated across the videos, these shadow effects are consistently transferred to novel views. 
Similarly, Fig. \ref{fig:reflection_shadow} (b) presents editing examples with realistic reflection effects, where surfaces reflect newly generated elements, such as flowers and trees. 
These reflections are reasonably propagated into novel views, maintaining consistency with the surrounding contents.

Both our model and the image editing method MagicQuill \cite{MagicQuill} benefit from being trained on real-world images and videos, which inherently contain realistic shadows and reflections. 
The training allows the models to handle such physical phenomenon with a certain level of accuracy.
However, it's important to note that the results, while visually plausible, do not strictly adhere to the physical laws of lighting and surface reflection. 
This opens up an interesting direction for future research, with potential solutions such as accurate 3D reconstruction integrated with physical simulation.

\section{limitation of camera motion}
Thanks to the 3D guidance and video diffusion model, our method handles input videos with significant view point changes.
However, it generates artifacts for camera motions involving 360-degree rotation. 
As shown in Fig. \ref{fig:failure_case}, while our method generates reasonable results in novel views, it introduces artifacts when the camera returns to a view similar to the edited image, such as distorted geometry (e.g. warped eyeglasses).
We attribute this limitation to the training data, which do not include videos with 360 degree rotations. 
A potential solution to this issue is augmenting the dataset with more diverse camera motions, particularly those involving full rotational views.

\section{Results on TU-Berlin dataset sketches}
\begin{figure*}[h]
    \centering
    \includegraphics[width=1.0\linewidth]{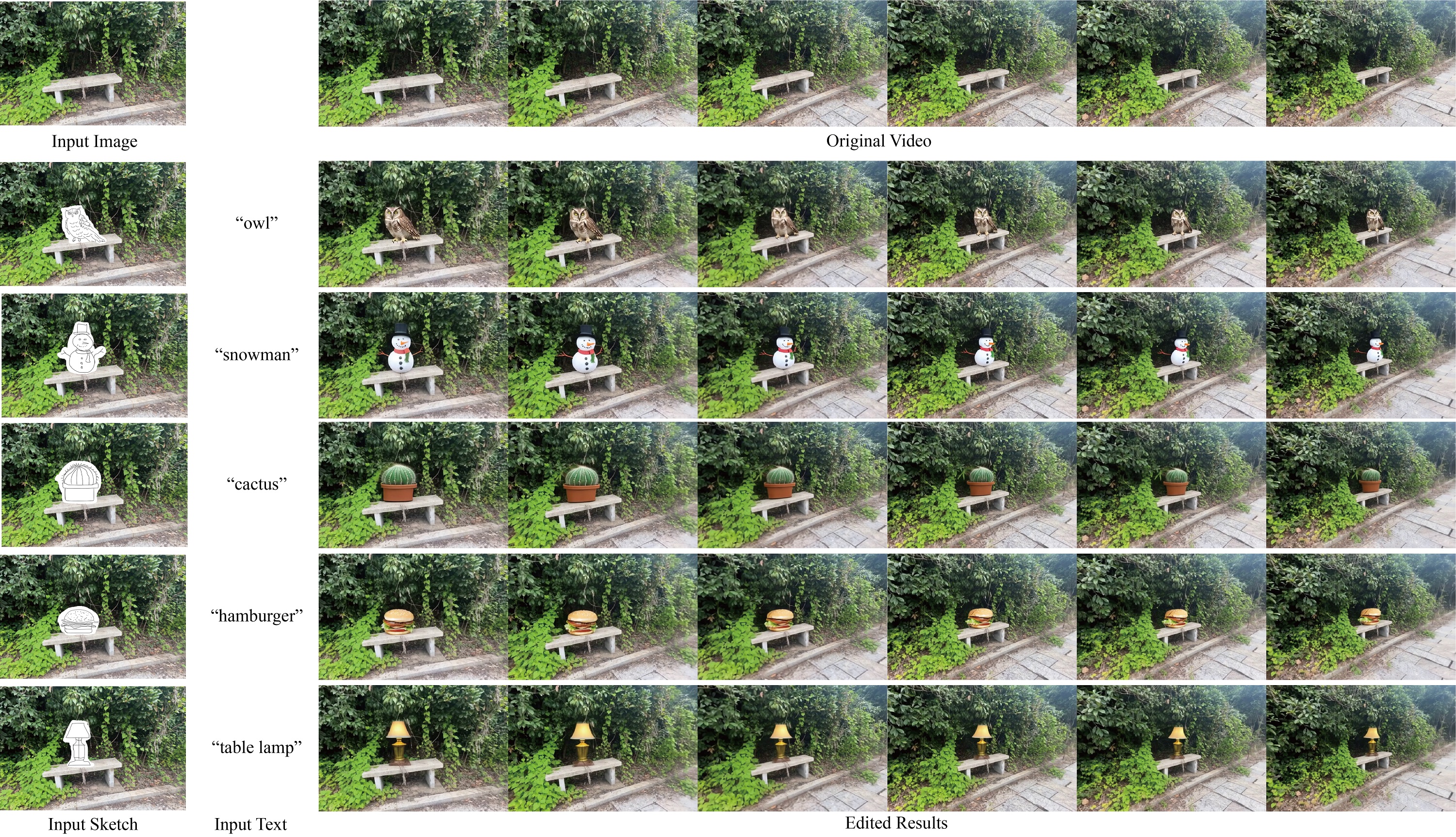}
    \caption{Editing results using sketches from various categories in the TU-Berlin dataset. On the left, the original images, texts, and sketches are shown, while the right side displays the original and edited video frames. Our method generates realistic video edits across a range of categories, including animals, plants, food, and daily objects. Original Video ©DL3DV-10K.
    }
    \Description{TU-Berlin}
    \label{fig:TU-Berlin}
\end{figure*}

Our method is robust for various categories of sketches. 
As shown in Fig. \ref{fig:TU-Berlin}, we apply sketches from the TU-Berlin dataset \cite{TU-Berlin}, which spans diverse categories such as animals, plants, foods and daily objects.
All of these sketches are successfully translated into realistic edited images, and the resulting content is effectively propagated into the videos.

\section{Robustness to size and position variations}
\begin{figure*}[h]
    \centering
    \includegraphics[width=1.0\linewidth]{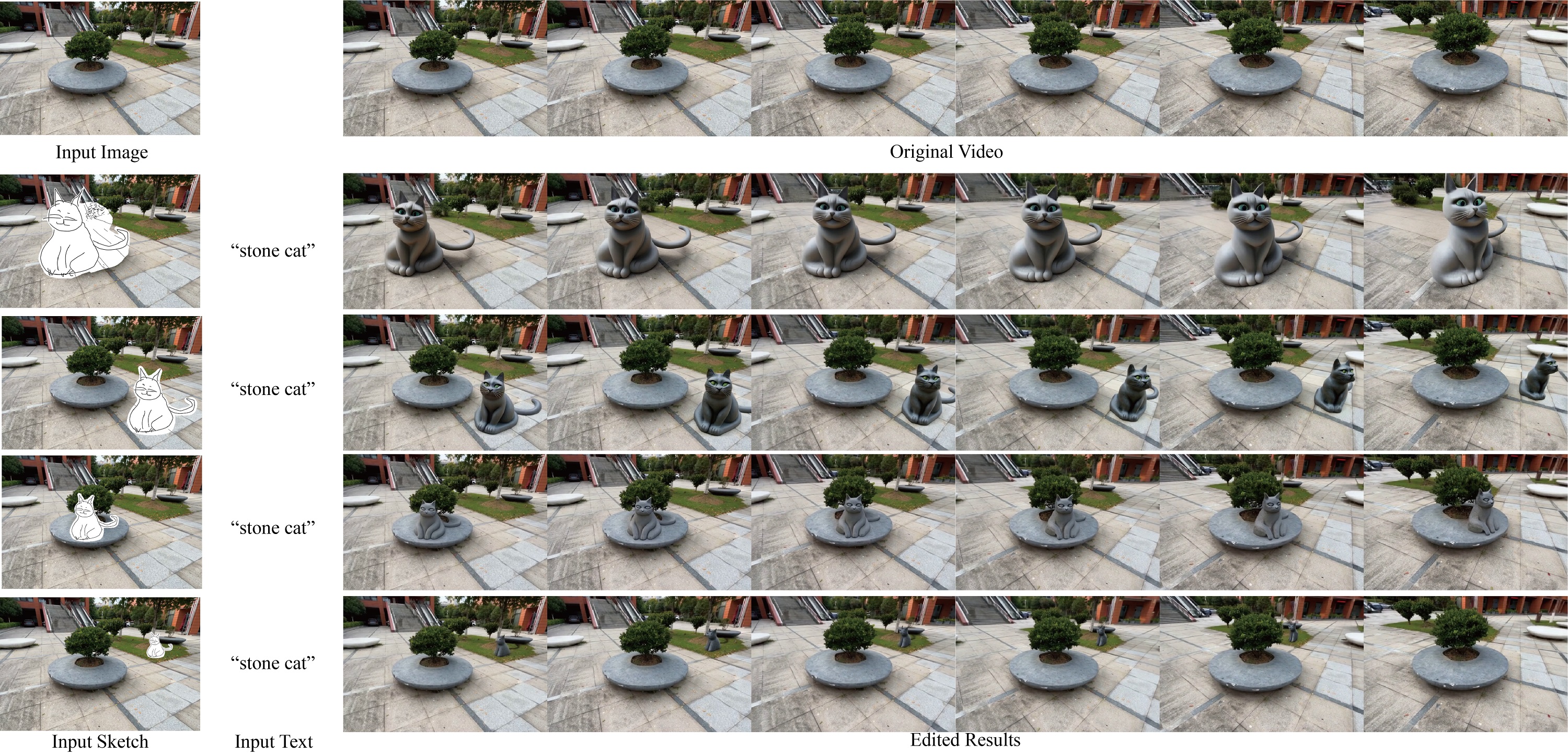}
    \caption{Editing results of the same sketch at varying scales and positions. Our method demonstrates robustness to changes in object size and placement, with realistic results across different positions. However, for very small objects (shown in the last row), the fine-grained details may become less sharp, leading to a slight loss of fidelity. Original Video ©DL3DV-10K.
    }
    \Description{size_position}
    \label{fig:size_position}
\end{figure*}

In Fig. \ref{fig:size_position}, we show the results of applying the same editing sketches at different positions and sizes. 
From the top to bottom, the size of the editing sketches decreases. 
Despite the size variation, all newly generated content maintains high realism and good quality. 
Furthermore, when the sketches are placed in different positions, they interact well with the surrounding environment as the view point changes. 
These examples demonstrate the robustness of our method. 
However, as shown in the last row in Fig. \ref{fig:size_position}, for extremely small objects, the fidelity to the sketches may degrade, particularly in fine-grained details. 
A potential solution to this issue could involve independently generating these regions in high resolution, followed with rescaling and fusion into original images.

\section{Robustness to sketch style variations}
\begin{figure*}[h]
    \centering
    \includegraphics[width=1.0\linewidth]{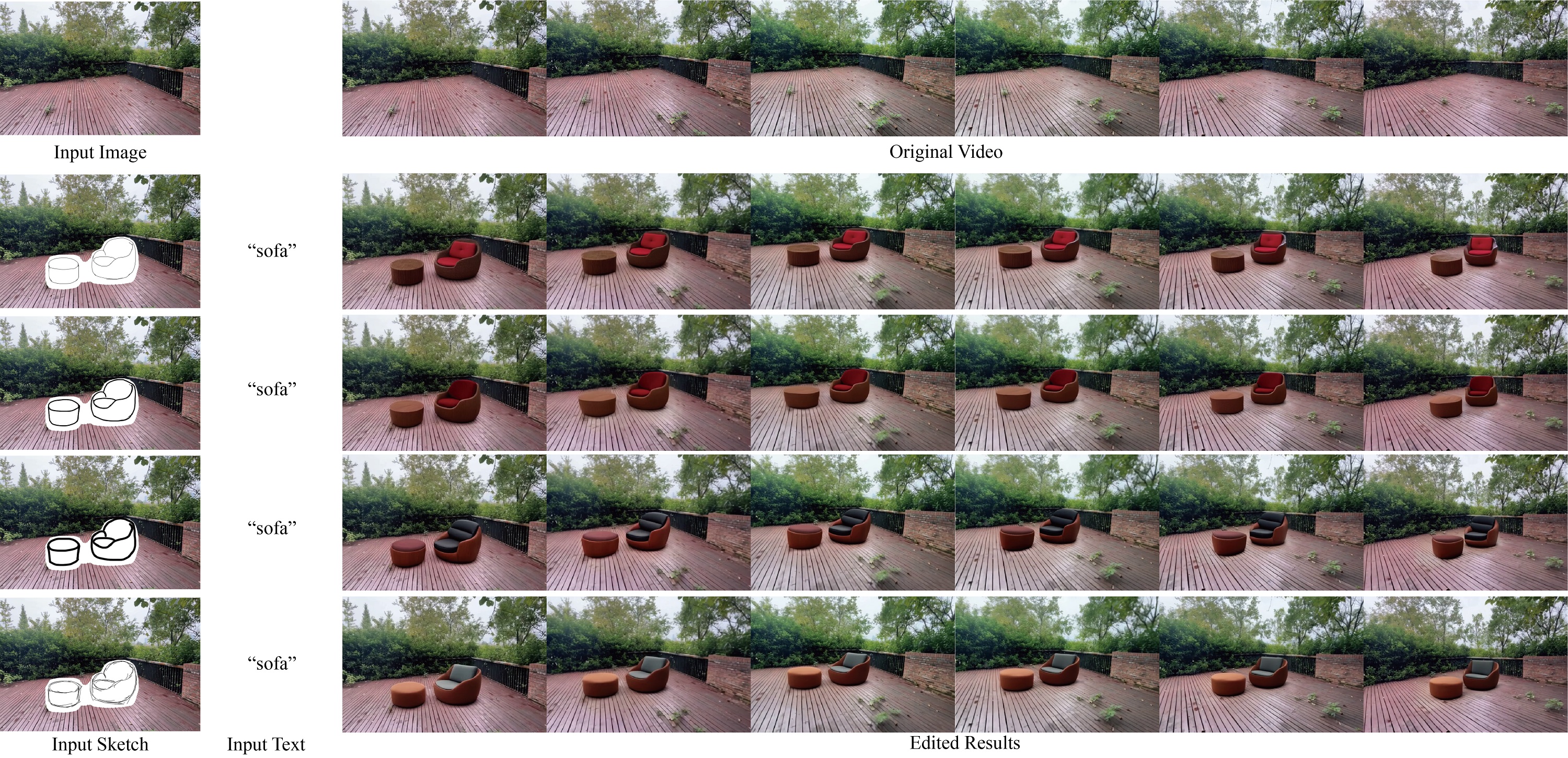}
    \caption{Editing results of the same sketch with different drawing styles. The image editing method, MagicQuill \cite{MagicQuill} effectively handles varying stroke widths and drawing styles. The edited operations are consistently propagated across video frames. Original Video ©DL3DV-10K.
    }
    \Description{style}
    \label{fig:style}
\end{figure*}

In Fig. \ref{fig:style}, we present results of applying the same editing sketches across various drawing styles. 
It can been seen that the image editing method MagicQuill \cite{MagicQuill} generates realistic edited images for different stroke widths and styles. 
The editing operations are effectively propagated across the video frames, demonstrating the robustness of our method in handling diverse sketch styles.

\section{More comparison Results}

\begin{figure*}[h]
    \centering
    \includegraphics[width=1.0\linewidth]{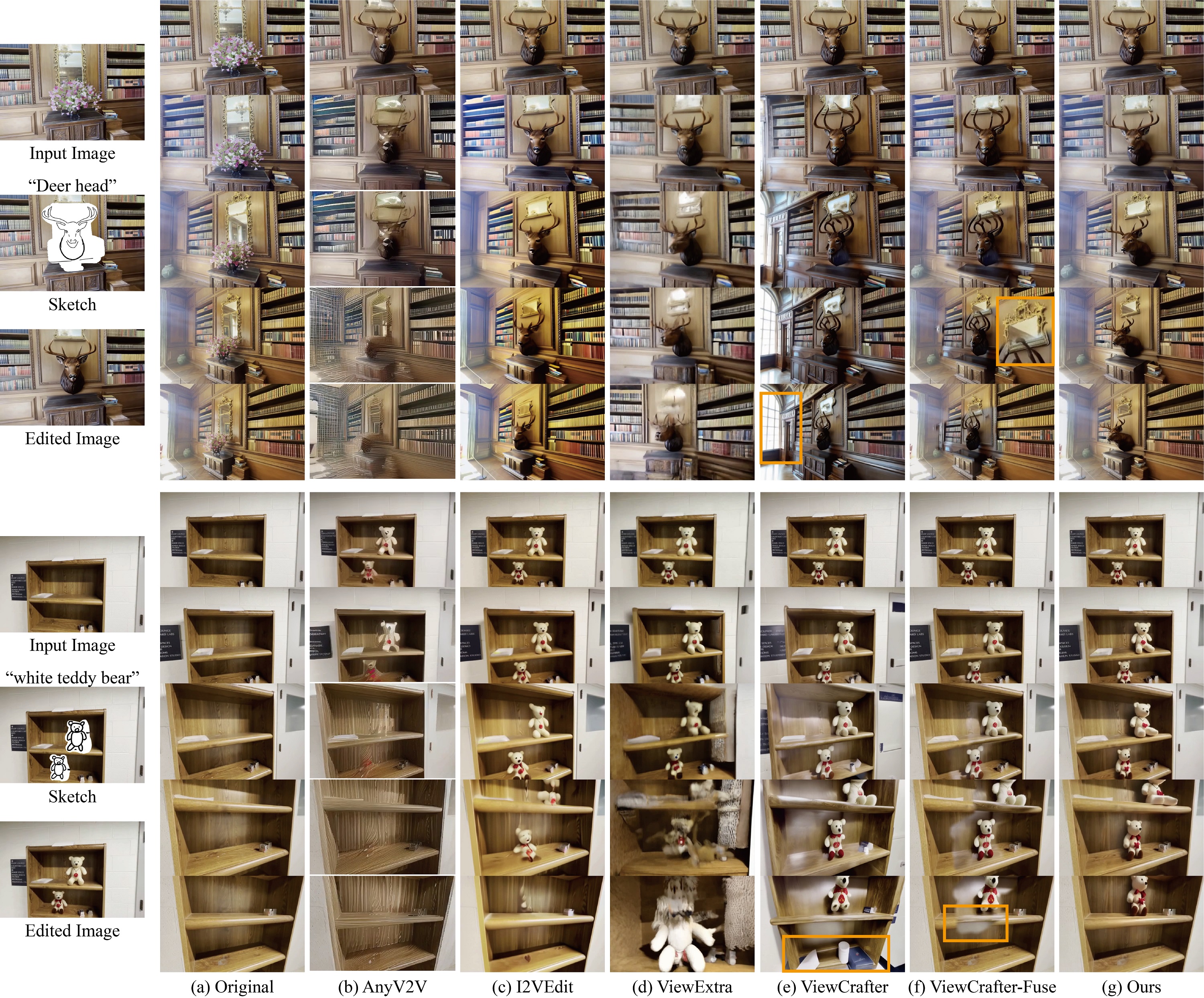}
    \caption{The comparison with existing methods, including AnyV2V \cite{AnyV2V}, I2VEdit \cite{I2VEdit}, ViewExtrapolation \cite{ViewExtrapolation} (shorted as ViewExtra), and ViewCrafter \cite{ViewCrafter}. 
    Unlike these approaches, which have noticeable artifacts in edited objects or inadvertently alter unedited regions when significantly changing view points, our method consistently produces the most realistic and coherent edited results. Original Video ©DL3DV-10K.
    }
    \Description{compare_supple}
    \label{fig:compare_supple}
\end{figure*}

\begin{figure*}[h]
    \centering
    \includegraphics[width=1.0\linewidth]{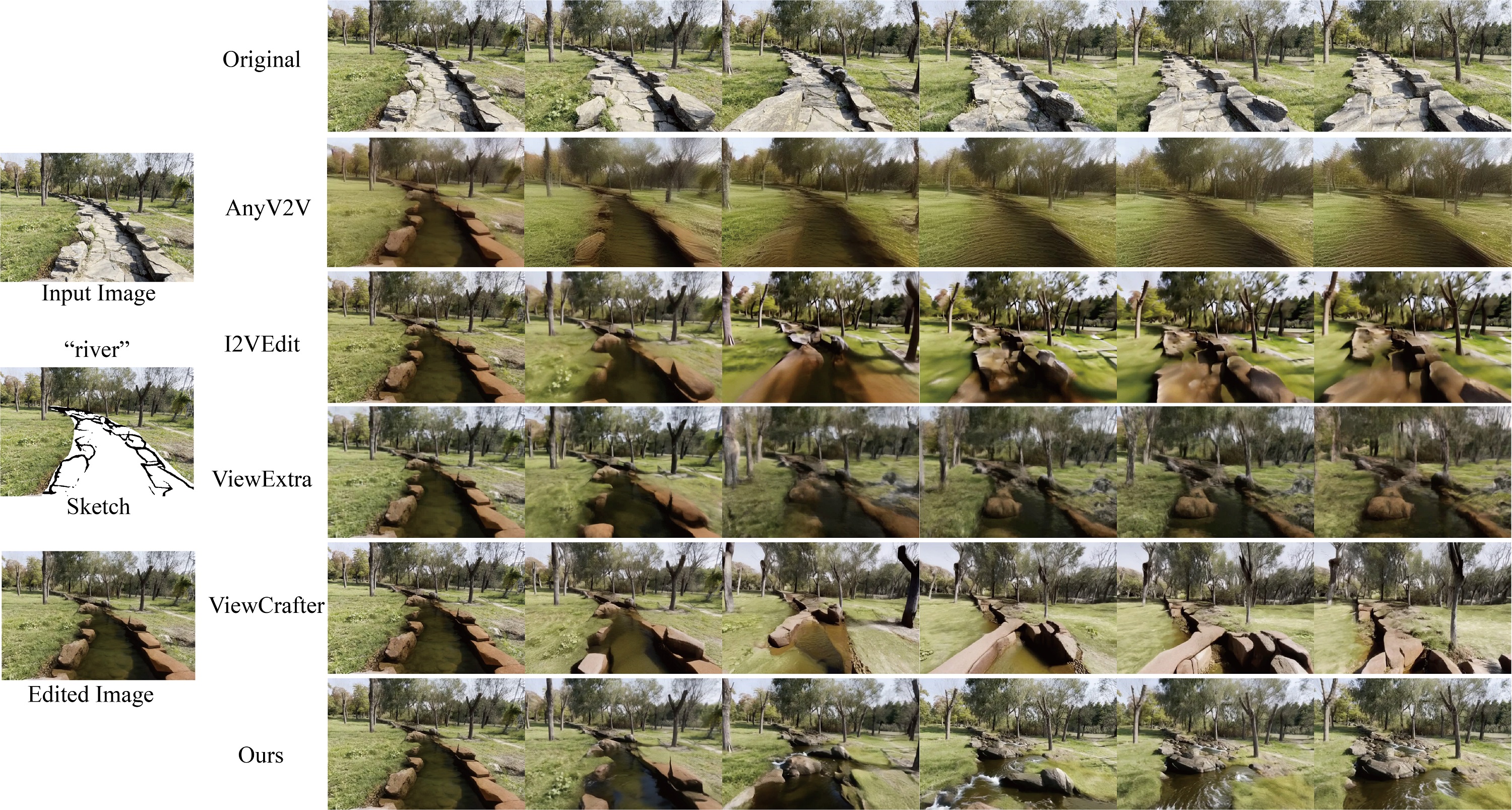}
    \caption{The comparison with existing methods, including AnyV2V \cite{AnyV2V}, I2VEdit \cite{I2VEdit}, ViewExtrapolation \cite{ViewExtrapolation} (shorted as ViewExtra), and ViewCrafter \cite{ViewCrafter}, for dynamic examples. 
    Our method generates realistic water flowing results, while other approaches generate fuzzy details or unreasonable transition. Original Video ©DL3DV-10K.
    }
    \Description{compare_dynamic_supple}
    \label{fig:compare_dynamic_supple}
\end{figure*}

In Fig. \ref{fig:compare_supple}, we show more comparison results with existing methods. Our method generates the most realistic and consistent video results.
In Fig. \ref{fig:compare_dynamic_supple}, we show comparison results for objects with slight dynamic motions. Our method outpeforms existing works.

\section{More application Results}

\begin{figure*}[h]
    \centering
    \includegraphics[width=1.0\linewidth]{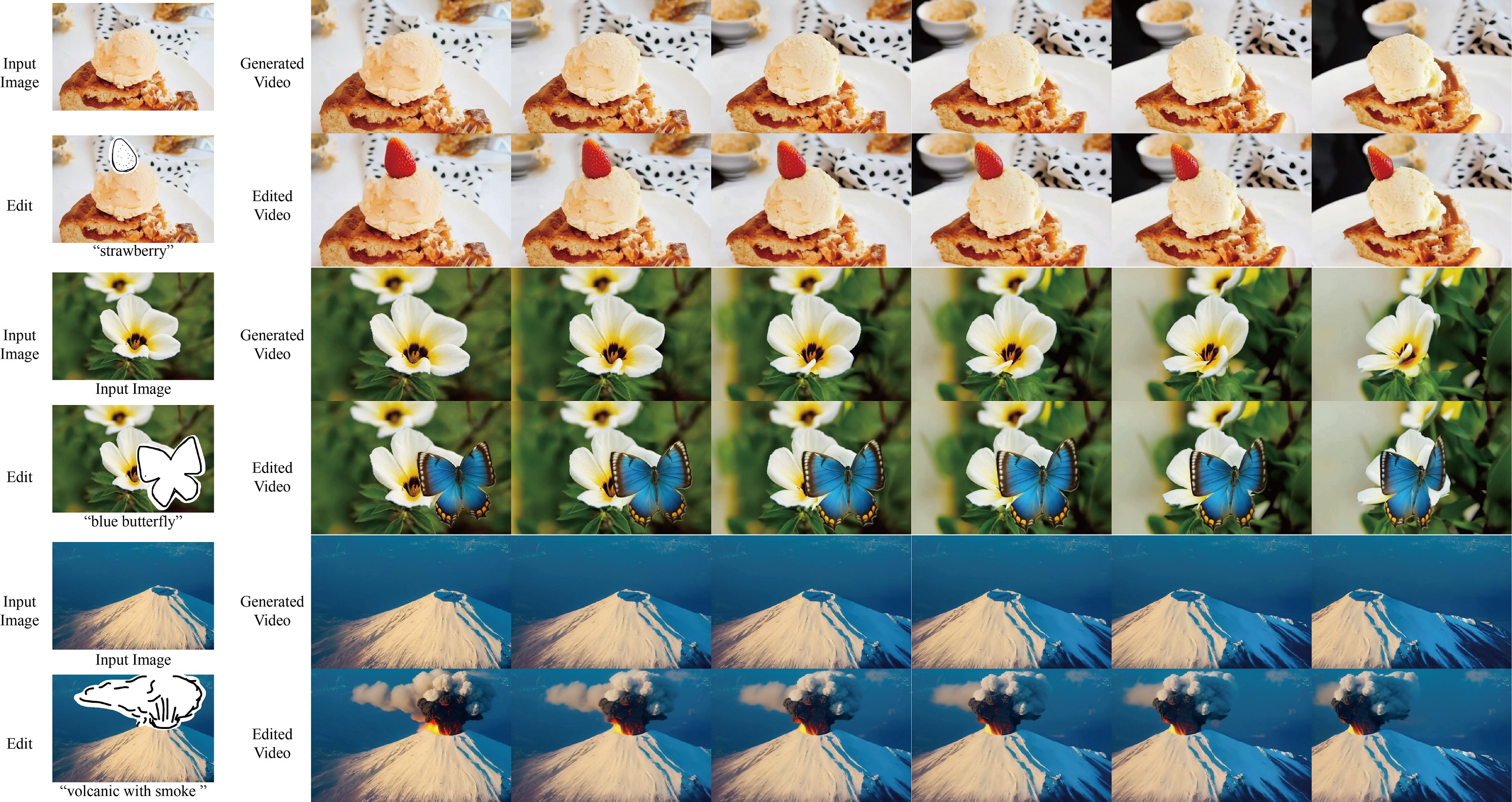}
    \caption{The application of image-to-video generation and editing. Given input images, users can control the camera trajectory to generate 3D scene videos. The local regions can further be edited based on sketches. Input Image ©Anastasia Belousova, ©Abdulla Nadeem, ©Tirachard Kumtanom.
    }
    \Description{application_3D_gen}
    \label{fig:application_3D_gen}
\end{figure*}

In Fig. \ref{fig:application_3D_gen}, we show more application results of image-to-video generation and editing results. Our method achieves interesting 3D scene video customization. 

\bibliographystyle{ACM-Reference-Format}
\bibliography{supple}